\batchmode
\makeatletter
\def\input@path{{\string"/Users/LLS/Dropbox/Research/Constructing good quantum bosonic codes via creating destructive interference/\string"}}
\makeatother
\documentclass[english,prl, two column, superscriptaddress]{revtex4-1}
\usepackage[T1]{fontenc}
\usepackage[latin9]{inputenc}
\usepackage{color}
\usepackage{babel}
\usepackage{array}
\usepackage{mathtools}
\usepackage{amsmath}
\usepackage{amssymb}
\usepackage{stackrel}
\usepackage{graphicx}
\usepackage{wasysym}
\usepackage[pdftex,unicode=true,pdfusetitle,
 bookmarks=true,bookmarksnumbered=false,bookmarksopen=false,
 breaklinks=false,pdfborder={0 0 0},pdfborderstyle={},backref=false,colorlinks=true]
 {hyperref}
\hypersetup{
 colorlinks,linkcolor=red,citecolor=blue}

\makeatletter

\providecommand{\tabularnewline}{\\}

\usepackage{braket}

\makeatother

\begin{document}
\global\long\def\half{\frac{1}{2}}%
\global\long\def\a{\alpha}%
\global\long\def\b{\beta}%
\global\long\def\g{\gamma}%
\global\long\def\c{\chi}%
\global\long\def\d{\delta}%
\global\long\def\o{\omega}%
\global\long\def\m{\mu}%
\global\long\def\s{\sigma}%
\global\long\def\n{\nu}%
\global\long\def\z{\zeta}%
\global\long\def\l{\lambda}%
\global\long\def\k{\kappa}%
\global\long\def\x{\chi}%
\global\long\def\r{\rho}%
\global\long\def\t{\theta}%
\global\long\def\G{\Gamma}%
\global\long\def\D{\Delta}%
\global\long\def\O{\Omega}%
\global\long\def\pr{\prime}%
\global\long\def\k{\kappa}%
\global\long\def\S{\mathcal{S}}%
\global\long\def\R{\mathcal{R}}%
\global\long\def\E{\mathcal{E}}%
\global\long\def\bra{\langle}%
\global\long\def\ket{\rangle}%
\global\long\def\dg{\dagger}%
\global\long\def\dgt{\ddagger}%
\global\long\def\tr{\text{Tr}}%
\global\long\def\Tr{\textsc{Tr}}%
\global\long\def\id{\mathcal{I}}%
\global\long\def\e{\bar{\eta}}%
\global\long\def\nb{\bar{n}}%
\global\long\def\ph{\hat{n}}%
\global\long\def\aa{\hat{a}}%
\global\long\def\pc{P_{\!\textnormal{\ensuremath{\mathtt{cat}}}}}%
\global\long\def\pl{P_{\!\textnormal{\ensuremath{\mathtt{gkp2}}}}}%
\global\long\def\pb{P_{\textnormal{\ensuremath{\mathtt{bin}}}}}%
\global\long\def\pg{P_{\!\textnormal{\ensuremath{\mathtt{gkp}}}}}%
\global\long\def\pn{P_{\!\textnormal{\ensuremath{\mathtt{num}}}}}%
\global\long\def\pp{P_{\!\textnormal{\ensuremath{\mathtt{code}}}}}%
\global\long\def\gkp{\textnormal{\ensuremath{\mathtt{gkp}}}}%
\global\long\def\cat{\textnormal{\ensuremath{\mathtt{cat}}}}%
\global\long\def\cc{\textnormal{\ensuremath{\mathtt{code}}}}%
\global\long\def\num{\textnormal{\ensuremath{\mathtt{num}}}}%
\global\long\def\bin{\textnormal{\ensuremath{\mathtt{bin}}}}%
\global\long\def\lat{\textnormal{\ensuremath{\mathtt{gkp2}}}}%
\global\long\def\xx{\hat{x}}%
\global\long\def\fe{F_{e}}%
\global\long\def\A{\mathcal{A}}%
\global\long\def\L{\mathcal{L}}%
\global\long\def\sab{\textnormal{\ensuremath{\mathtt{sab}}}}%
\global\long\def\okb{\textnormal{\ensuremath{\mathtt{okb}}}}%
\global\long\def\sac{\textnormal{\ensuremath{\mathtt{sac}}}}%
\global\long\def\st{\textnormal{\ensuremath{\mathtt{stn^{\pr}}}}}%
\global\long\def\ot{\otimes}%
\global\long\def\er{\mathbb{E}}%
\global\long\def\sh{\textnormal{\ensuremath{\mathtt{shor}}}}%
\global\long\def\shp{\textnormal{\ensuremath{\mathtt{shor'}}}}%
\global\long\def\shpp{\textnormal{\ensuremath{\mathtt{shor''}}}}%
\global\long\def\lket#1{\left|#1\right\rangle }%

\title{Designing good bosonic quantum codes via creating destructive interference}
\author{Linshu Li }
\affiliation{Yale Quantum Institute, Departments of Applied Physics and Physics,
Yale University, New Haven, CT 06511, USA}
\author{Dylan J. Young}
\affiliation{Yale Quantum Institute, Departments of Applied Physics and Physics,
Yale University, New Haven, CT 06511, USA}
\author{Victor V. Albert}
\affiliation{Yale Quantum Institute, Departments of Applied Physics and Physics,
Yale University, New Haven, CT 06511, USA}
\affiliation{Walter Burke Institute for Theoretical Physics and Institute for Quantum
Information and Matter, California Institute of Technology, Pasadena,
California 91125, USA}
\author{Kyungjoo Noh}
\affiliation{Yale Quantum Institute, Departments of Applied Physics and Physics,
Yale University, New Haven, CT 06511, USA}
\author{Chang-Ling Zou}
\affiliation{Yale Quantum Institute, Departments of Applied Physics and Physics,
Yale University, New Haven, CT 06511, USA}
\affiliation{Key Laboratory of Quantum Information, University of Science and Technology
of China, Hefei, Anhui 230026, China}
\author{Liang Jiang}
\affiliation{Yale Quantum Institute, Departments of Applied Physics and Physics,
Yale University, New Haven, CT 06511, USA}
\begin{abstract}
Continuous-variable systems protected by bosonic quantum error-correcting
codes have emerged as a promising platform for quantum information
processing. To date, design of codewords has centered on optimizing
the occupation of basis states in the error-relevant basis. Here,
we propose utilizing the phase degree of freedom in basis state probability
amplitudes to devise codes that feature destructive interference,
and thus reduced overlap, between error codewords. To showcase, we
first consider the correction of excitation loss using single-mode
codes with Fock-space parity structure and show that, with a tailored
``two-level'' recovery, altering the signs of probability amplitudes
can significantly suppress decoherence. We then study the joint channel
of excitation loss and Kerr effect, and show the critical role of
nontrivial phase for optimal quantum codes for such channels. The
principle is extended to improve bosonic codes defined in other bases
and multi-qubit codes, showing its broader applicability in quantum
error correction.
\end{abstract}
\maketitle

\paragraph*{Introduction}

Quantum operations with continuous variables represent a promising
alternative path towards scalable quantum computing and communication
\cite{Knill2001a,Braunstein2005,Gottesman2001,Ralph2003}. Similar
to qubit-based systems, a major challenge for faithful bosonic quantum
information is to store, manipulate and communicate the encoded information
in the presence of noise, such as excitation loss (aka amplitude damping,
pure loss), phase-space drift, dephasing and cavity nonlinearities.
To overcome excitation loss that fundamentally limits the cavity lifetime,
multi-mode codes were first introduced \cite{Chuang1997,Leung1997,Ouyang2014,Bergmann2016a,Niu2018};
for phase-space drift, Gottesman, Kitaev, and Preskill ($\gkp$) codes
\cite{Gottesman2001,Harrington2001} were proposed. More recently,
motivated by the potential to utilize higher excitation states in
a bosonic Hilbert space and perform hardware-efficient operations,
single-mode codes for excitation loss such as $\cat$ codes \cite{Leghtas2013,Mirrahimi2014,Ofek2016,Bergmann2016,Li2017}
and binomial ($\bin$) codes \cite{Michael2016} were developed. Meanwhile,
progress in superconducting circuit quantum electrodynamics (QED)
\cite{Sun2014,Krastanov2015,Heeres2015} -- e.g. FPGA adaptive control
\cite{Ofek2016}, readout of excitation parity \cite{Sun2014} and
universal control of cavity states \cite{Krastanov2015,Heeres2015}
-- has opened up possibilities once thought unreachable, including
implementing arbitrary quantum channels on a bosonic mode \cite{Shen2017,Lloyd2001}.
With the advances, error-corrected $\cat$ and $\bin$ qubits and
the associated universal quantum gate sets have been demonstrated,
respectively \cite{Ofek2016,Heeres2017,Hu2018}. These capabilities
are essential for higher-level operations such as distributing error-corrected
entangled states \cite{Axline2018} and quantum gate teleportation
\cite{Chou2018}.

For the aforementioned codes, in the computational basis relevant
to the error under consideration, i.e. Fock basis for $\cat/\bin$
codes and position/momentum basis for square-lattice $\gkp$ codes,
the codewords are spanned by distinct subsets of basis states with
positive probability amplitudes. Recently, $\gkp$ codes have been
found to protect against excitation loss extremely well, even though
they were not originally designed for loss errors \cite{Albert2018,Noh2018}.
Similar to the newly discovered numerically optimized codes \cite{Michael2016,Albert2018},
when expressed in the Fock basis, their codewords do not possess parity
structure, yet feature negative probability amplitudes. The findings
inspire us to better understand the recipe behind desirable quantum
error-correction (QEC) capabilities and generalize the findings to
devise more efficient codes.

In this Letter, we explore the conjugate degree of freedom to basis
state occupation -- the relative phases -- and demonstrate its critical
role for efficient quantum codes. Firstly, we show that tuning the
phase degree of freedom in the codewords can improve code performance
via making error codewords interfere destructively. As illustrations,
concerning the correction of excitation loss, we modify $\bin$ and
$\cat$ codes by periodically altering the signs of probability amplitudes
in one codeword, and show that the sign alteration (SA) effectively
reduces the overlap between selected error codewords. The periodic
SA can be experimentally realized via adding a cavity Kerr to the
encoding procedure. To capture the enhanced separation created by
SA, we propose a two-level recovery that yields error protection close
to optimal at loss rates of practical interest. The enhancement naturally
leads to a question: Is phase degree of freedom necessary for $\mathit{optimal}$
code constructs? Using biconvex optimization of encoding and decoding
\cite{Reimpell2005,Kosut2009,Noh2018}, we compute the optimal code
given an error process and find that, although nontrivial phase (neither
$0$ nor $\pi$) is redundant for excitation loss, it is necessary
for more complicated ones such as the joint channel of excitation
loss and Kerr effect.

Extending the principle to codes defined in bases beyond Fock, we
note that GKP codes over non-square lattices, which better overcome
phase-space drift than square lattice GKP, can be generated from the
later through SA in position-momentum basis \cite{Gottesman2001}.
In addition, noting that well-known multi-qubit CSS codes \cite{Shor1995,Steane1996,Kitaev2003}
share the same feature of all positive probability amplitudes with
$\bin$ and $\cat$ codes, we modify the nine-qubit Shor code for
enhanced protection against Pauli errors and amplitude damping error,
demonstrating the wide applicability of the principle for enhanced
quantum error correction.

\paragraph{Principle}

\begin{figure}[t]
\centering{}\includegraphics[scale=0.41]{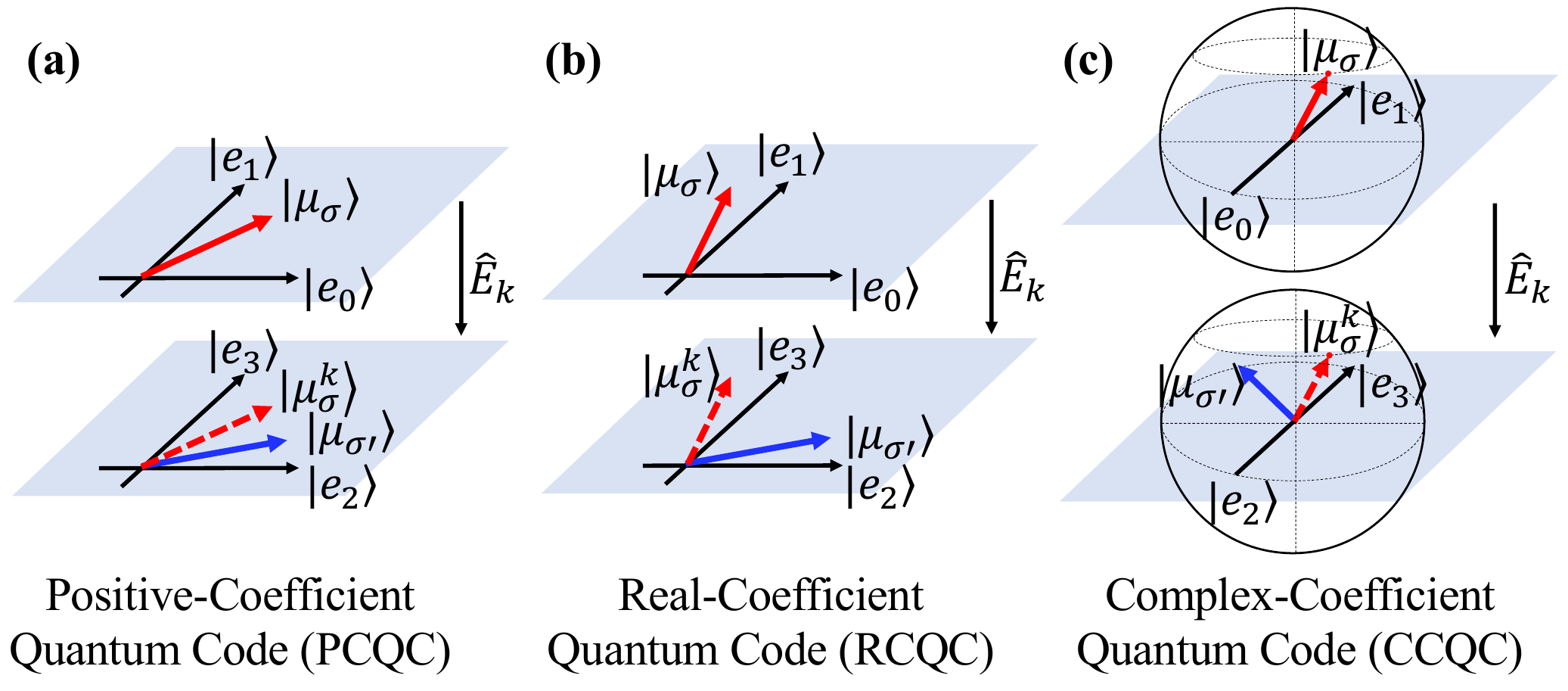}\caption{Illustrations of error codeword overlap for (a) positive-coefficient,
(b) real-coefficient and (c) complex-coefficient quantum code. Here,
$|\mu_{\sigma}\protect\ket$ are codewords and $|e_{i}\protect\ket$
are computational basis states that span $|\mu_{\sigma}\protect\ket$.
By allowing (b) $\theta_{n}^{\sigma}=0,\,\pi$ and (c) $\theta_{n}^{\sigma}\in[0,2\pi)$,
after the occurrence of $\hat{E}_{k}$ to $|\mu_{\sigma}\protect\ket$
(red solid vector), error codeword $|\mu_{\sigma}^{k}\protect\ket$
(red dashed vector), spanned by the same $S_{\protect\s'}$ as another
codeword $|\mu_{\sigma'}\protect\ket$ (blue vector), remains largely
separable from $|\mu_{\sigma'}\protect\ket$ due to destructive interference.
For clarity, we assume each codeword consists of two basis states.
\label{fig:Graphical illustration of destructive interference} }
\end{figure}
We begin with introducing the principle that engineering the phase
degree of freedom in codewords can create destructive interference
between them after error, thus suppressing undesired overlap. Given
a qudit code embedded in a Hilbert space $\mathcal{H}=\{|n\ket\}$
where $n$ indexes computational basis states, normalized codewords
in a chosen logical basis can be written as
\begin{equation}
|\mu_{\sigma}\ket=\sum_{n\in S_{\s}}c_{n}^{\sigma}e^{i\theta_{n}^{\sigma}}|n\ket\label{eq:abstract quantum code}
\end{equation}
where $\sigma=0,1,...,d-1$ labels codewords, $c_{n}^{\sigma}>0$
and $S_{\s}$ is a subset of the computational basis chosen for each
codeword. The error process that the code in Eq.~(\ref{eq:abstract quantum code})
is devised to protect against can be expressed in the Kraus representation
$\mathcal{E}(\rho)=\sum_{k}\hat{E}_{k}\rho\hat{E}_{k}^{\dagger}$,
with each operator $\hat{E}_{k}$ associated with an error event.
The decoherence under $\mathcal{E}$ is captured by the QEC matrix
$M_{kl,\sigma\sigma'}=\bra\mu_{\sigma}|\hat{E}_{k}^{\dagger}\hat{E}_{l}|\mu_{\sigma'}\ket$
\cite{Bennett1996,Knill1997}, i.e. overlap between codewords under
errors, and
\begin{align}
\bra\mu_{\sigma}|\hat{E}_{k}^{\dagger}\hat{E}_{l}|\mu_{\sigma'}\ket & =\stackrel[m,n]{+\infty}{\sum}c_{m}^{\sigma}c_{n}^{\sigma'}e^{i(\theta_{n}^{\sigma'}-\theta_{m}^{\sigma})}\bra m|\hat{E}_{k}^{\dagger}\hat{E}_{l}|n\ket\,.\label{eq:bit-flip QEC terms}
\end{align}
We see that, depending on the nature of error and its relation to
the computational basis, magnitude $c_{n}^{\sigma}$ and phase $\theta_{n}^{\sigma}$
are both critical in suppressing Eq.~(\ref{eq:bit-flip QEC terms}).

Figure \ref{fig:Graphical illustration of destructive interference}
illustrates how phase allows codewords, after an error occurs, to
remain distinguishable. Fig.~\ref{fig:Graphical illustration of destructive interference}(a)
shows the case of \textit{positive-coefficient quantum code} (PCQC),
for which $\theta_{n}^{\sigma}=0$ and it relies only on $c_{n}^{\sigma}$
and a wise choice of $S_{\s}$ to ensure orthogonality between codewords
and QEC capacity. Examples include $\bin$, $\cat$, multi-mode \cite{Chuang1997,Leung1997},
square-lattice $\gkp$ codes for phase-space drift and multi-qubit
CSS codes. Once error $\hat{E}_{k}$ takes $|\mu_{\sigma}\ket$ to
the subspace $|\mu_{\sigma'}\ket$ lies in, $|\mu_{\sigma'}\ket$
and error word $|\mu_{\sigma}^{k}\ket\coloneqq\hat{E}_{k}|\mu_{\sigma}\ket/\sqrt{\bra\mu_{\sigma}|\hat{E}_{k}^{\dagger}\hat{E}_{k}|\mu_{\sigma}\ket}$
will overlap, inducing decoherence. However, if we consider a \textit{real-coefficient
quantum code} (RCQC) where $\theta_{n}^{\sigma}\in\{0,\,\pi\}$ {[}Fig.~\ref{fig:Graphical illustration of destructive interference}(b){]},
or even \textit{complex-coefficient quantum code} (CCQC) where $\theta_{n}^{\sigma}\in[0,\,2\pi)$
{[}Fig.~\ref{fig:Graphical illustration of destructive interference}(c){]},
dependent on the error under consideration and selected computational
basis, larger separation between $|\mu_{\sigma}^{k}\ket$ and $|\mu_{\sigma'}\ket$
can potentially be realized due to destructive interference between
basis states. Without losing generality, we focus on quantum codes
that encode a qubit.

To see how better codes can emerge as a result, we begin with a neat
example -- the bosonic $\sqrt{17}$-code \cite{Michael2016} --
that corrects a single loss without a parity structure key to other
codes
\begin{eqnarray}
|0_{L}\ket & = & \frac{1}{\sqrt{6}}(\sqrt{7-\sqrt{17}}|0\ket+\sqrt{\sqrt{17}-1}|3\ket)\,,\label{eq:sqrt17-1}\\
|1_{L}\ket & = & \frac{1}{\sqrt{6}}(\sqrt{9-\sqrt{17}}|1\ket-\sqrt{\sqrt{17}-3}|4\ket)\,.\label{eq: sqrt17-2}
\end{eqnarray}
$\hat{a}|0_{L}\ket$ ($\hat{a}$ is annihilation operator) and $|1_{L}\ket$,
spanned by different Fock states, do not overlap. Meanwhile, one can
test that, due to \textit{destructive interference,} $\hat{a}|1_{L}\ket\propto\sqrt{9-\sqrt{17}}|0\ket-2\sqrt{\sqrt{17}-3}|3\ket$
is also orthogonal to $|0_{L}\ket$, thus allowing the code to fully
correct one excitation loss.

\paragraph{Sign-altered $\protect\bin$ code for excitation loss}

The energy decay of an oscillator is described by excitation loss
channel $\mathcal{N}_{\gamma}$ ($\gamma$ is loss rate), for which
Kraus operator $\hat{E}_{k}=\sqrt{\frac{\gamma^{k}}{k!}}(1-\gamma)^{\frac{\hat{n}}{2}}\hat{a}^{k}$
\cite{Chuang1997,Michael2016,Li2017} ($\hat{n}=\hat{a}^{\dagger}\hat{a}$
is the excitation number operator) is associated with losing $k$
excitations. To correct the multi-loss events, based on $\bin(N,S)$
that corrects exactly $S-1$ losses for $N\geq S$
\begin{eqnarray}
|0_{\mathrm{\bin}}/1_{\mathrm{\bin}}\ket & = & 2^{-\frac{N-1}{2}}\stackrel[\mathrm{p}\,\mathrm{even/odd}]{[0,N]}{\sum}\sqrt{\left(\begin{array}{c}
N\\
p
\end{array}\right)}|pS\ket\,,\label{eq:orginal binomial code}
\end{eqnarray}
we apply a periodic SA to $|0_{\mathrm{\bin}}\ket$ while keeping
those of $|1_{\mathrm{\bin}}\ket$ unchanged, and obtain sign-altered
binomial ($\sab$) code with
\begin{eqnarray}
|0_{\sab}\ket & = & 2^{-\frac{N-1}{2}}\stackrel[\mathrm{p}\,\mathrm{even}]{[0,N]}{\sum}(-1)^{\frac{p}{2}}\sqrt{\left(\begin{array}{c}
N\\
p
\end{array}\right)}|pS\ket\,,\label{eq:SA-Bin 0}
\end{eqnarray}
and $|1_{\sab}\ket=|1_{\mathrm{\bin}}\ket$. With the same parity
structure, $\sab(N,S)$ code also corrects $S-1$ losses perfectly
as $\bin(N,S)$. The advantage emerges when we examine higher-order
QEC matrix entries $\bra\mu_{\sigma}|\hat{E}_{k}^{\dagger}\hat{E}_{S+k}|\mu_{\sigma'}\ket\ (k=0,1,\ldots)$:
Here, the $(-1)^{\frac{p}{2}}$ in Eq.~(\ref{eq:SA-Bin 0}) leads
to destructive interference between the two error codewords, reducing
the overlap responsible for logical-X error.

The encoding procedure of $\sab$ code builds on that of $\bin$ code
with an additional Kerr unitary that imprints SA to $|0_{\mathrm{\bin}}\ket$
while acting trivially on $|1_{\mathrm{\bin}}\ket$ as a result of
the parity structure. To see this, we first introduce Kerr unitary
$\hat{U}_{\mathrm{Kr}}=e^{\frac{1}{2}iKt\hat{n}^{2}}$ with strength
coefficient $K$. Denoting $\hat{U}_{\mathrm{S}}=\exp[i\pi\hat{n}^{2}/(2S)^{2}]$,
we see that
\begin{eqnarray}
\hat{U}_{\mathrm{S}}|0_{\mathrm{\bin}}\ket=|0_{\mathrm{\sab}}\ket,\ \hat{U}_{\mathrm{S}}|1_{\mathrm{\bin}}\ket & = & e^{i\frac{\pi}{4}}|1_{\mathrm{\sab}}\ket\,.\label{eq:kerr applied to encoded bin}
\end{eqnarray}
Note that $e^{i\frac{\pi}{4}}$ can be removed by redefining $|1_{\mathrm{\sab}}\ket$.
Hence the encoding of $\sab$ is realized as $\mathcal{S}_{\sab}(\cdot)=\hat{U}_{\mathrm{S}}\mathcal{S}_{\bin}(\cdot)\hat{U}_{\mathrm{S}}^{\dagger}$.

\paragraph{Two-level recovery}

A complete QEC process is described by the effective qubit channel
$\mathcal{E}=\mathcal{S}^{-1}\circ\mathcal{R}\circ\mathcal{N}\circ\mathcal{S}$
that consists of encoding $\mathcal{S}$, error channel $\mathcal{N}$,
recovery $\mathcal{R}$ and decoding $\mathcal{S}^{-1}$ \cite{Albert2018}.
Given $\mathcal{S}$ and $\mathcal{N}$, the code performance then
depends on choice of $\mathcal{R}$ -- in the case of $\sab$, we
need to design an $\mathcal{R}$ that effectively captures the enhanced
separation between error codewords.

We first recall the recovery proposed for equally-spaced codes with
spacing $S$ \cite{Sun2014,Michael2016,Li2017}. The recovery $\mathcal{R}^{(1)}=\{\hat{R}_{0}^{(1)},\hat{R}_{1}^{(1)},\ldots,\hat{R}_{S-1}^{(1)}\}$
and Kraus operator $\hat{R}_{i}^{(1)}=\hat{U}_{i}^{(1)}\hat{\Pi}_{i\bmod S}$
where $\hat{\Pi}_{i\bmod S}$ is the projection operator into the
subspace with excitation number $i$ modulo $S$, and unitary $\hat{U}_{i}^{(1)}$
performs state transfer $|\mu_{\sigma}^{(S-i)\bmod S}\ket\leftrightarrow|\mu_{\sigma}\ket$.
$\mathcal{R}^{(1)}$ makes use of the parity structure to correct
the first $S-1$ excitation losses, and losses beyond the first $S-1$
will lead to bit-flip error. As such, here we call it \textquotedblleft one-level\textquotedblright{}
recovery.

To capture the enhanced error codeword separation, we propose a new
recovery $\mathcal{R}^{(2)}$ that, in addition to correcting the
first $S-1$ losses, exploits the component in error word $|\mu_{\bar{\sigma}}^{S+k}\ket$
that is orthogonal to $|\mu_{\sigma}^{k}\ket$ ($k=0,1,...,S-1$),
i.e. $|\mu_{\bar{\sigma}}^{S+k}\ket-\bra\mu_{\sigma}^{k}|\mu_{\bar{\sigma}}^{S+k}\ket|\mu_{\sigma}^{k}\ket$.
Since events with $S$ to $2S-1$ losses are also partially corrected,
we call $\mathcal{R}^{(2)}$ ``two-level'' recovery \cite{Albert2018}.
We note that $\mathcal{R}^{(2)}$ improves the performance of $\bin$
compared to $\mathcal{R}^{(1)}$ \cite{Albert2018}, yet the enhancement
is much more pronounced for $\sab$ due to the intentionally enlarged
separation between $|\mu_{\bar{\sigma}}^{S+k}\ket$ and $|\mu_{\sigma}^{k}\ket$. 

The $1^{\mathrm{st}}$ level of $\mathcal{R}^{(2)}$, similarly to
$\mathcal{R}^{(1)}$ \cite{Michael2016}, fully corrects the first
$S-1$ losses. Each Kraus operator consists of a projection and a
restoring unitary 
\begin{eqnarray}
\hat{R}_{k}^{(2)} & = & \underset{\sigma}{\sum}(|\mu_{\sigma}\ket\bra\mu_{\sigma}^{k}|+\hat{U}_{k}^{\mathrm{res}})\hat{P}_{k}\,.
\end{eqnarray}
Here, $k=0,1,\ldots,S-1$, $\hat{P}_{k}=\sum_{\sigma}|\mu_{\sigma}^{k}\ket\bra\mu_{\sigma}^{k}|$
projects to each error subspace, and $\hat{U}_{k}^{\mathrm{res}}$
finishes the unitary rotation in $\mathrm{Span}\{|\mu_{\sigma}\ket,|\mu_{\sigma}^{k}\ket\}$
-- for $k\neq0$, it is simply $|\mu_{\sigma}^{k}\ket\bra\mu_{\sigma}|$.

Also, $\mathcal{R}^{(2)}$ has a $2^{\mathrm{nd}}$ level with $S$
Kraus operators
\begin{equation}
\hat{R}_{S+k}^{(2)}=\underset{\sigma}{\sum}(|\mu_{\bar{\sigma}}\ket\bra\nu_{\bar{\sigma}}^{k}|+\hat{U}_{S+k}^{\mathrm{res}})\hat{P}_{S+k}
\end{equation}
where, for $k=0,1,\ldots,S-1$, the normalized $|\nu_{\bar{\sigma}}^{k}\ket\propto|\mu_{\bar{\sigma}}^{S+k}\ket-\bra\mu_{\sigma}^{k}|\mu_{\bar{\sigma}}^{S+k}\ket|\mu_{\sigma}^{k}\ket$
is recoverable and $\hat{P}_{S+k}=\sum_{\sigma}|\nu_{\bar{\sigma}}^{k}\ket\bra\nu_{\bar{\sigma}}^{k}|$.
To make $\mathcal{R}^{(2)}$ a CPTP map, we add $\hat{R}_{2S+1}^{(2)}=\hat{V}_{\mathrm{res}}(\hat{I}_{\mathcal{H}}-\sum_{i=0}^{2S}\hat{P}_{i})$
where $\hat{I}_{\mathcal{H}}$ is the identity operator on the entire
Hilbert space and $\hat{V}_{\mathrm{res}}$ an arbitrary unitary acting
on the complementary subspace of $\{|\mu_{\sigma}\ket\}\cup\{|\mu_{\sigma}^{k}\ket\}\cup\{|\nu_{\bar{\sigma}}^{k}\ket\}$.

To quantify the performance of $\mathcal{E}$, we use channel fidelity
(aka. entanglement fidelity) defined as 
\begin{eqnarray}
F & \coloneqq & \bra\Psi|\mathcal{I}_{A}\otimes\mathcal{E}_{B}(|\Psi\ket\bra\Psi|)|\Psi\ket\,,\label{eq:channel fidelity}
\end{eqnarray}
where $\mathcal{I}_{A}$ is an identity channel on qubit A and $|\Psi\ket=\frac{1}{\sqrt{2}}(|0_{A}0_{B}\ket+|1_{A}1_{B}\ket)$
is a maximally entangled state of qubits A and B. 
\begin{figure}[tb]
\centering{}\includegraphics[scale=0.37]{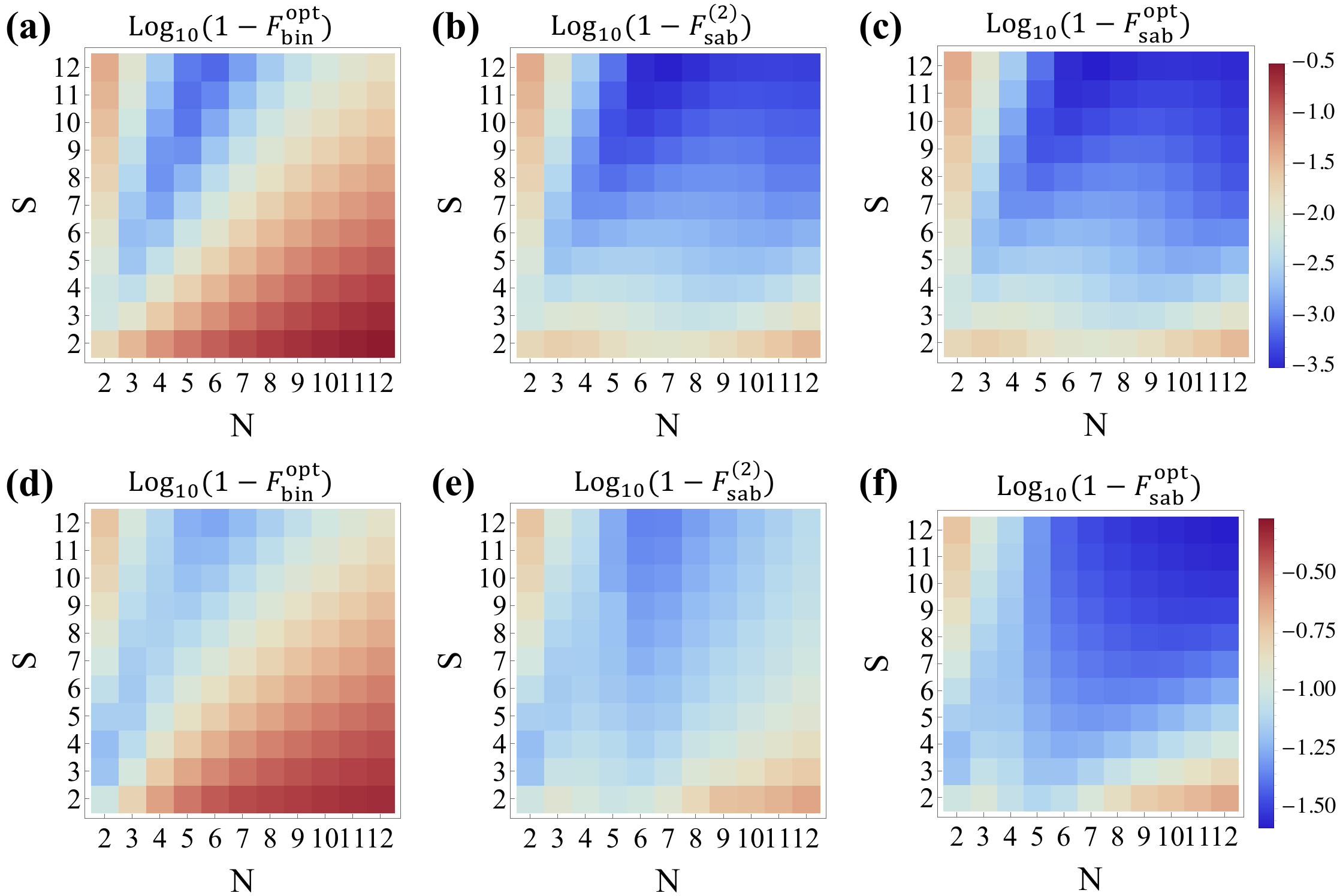}\caption{Channel infidelities (in logarithmic scale) for (a-c) $\protect\bin$
code with optimal recovery, $\protect\sab$ code with two-level recovery
and $\protect\sab$ code with optimal recovery, respectively, at $\gamma=0.1$,
and (d-f) same as (a-c) except for $\gamma=0.25$. Each point represents
a code with associated $S$ and $N$.\label{fig:optimal infidelities for bin and sab}}
\end{figure}
In Fig.~\ref{fig:optimal infidelities for bin and sab}, we present
the performance of $\mathcal{R}^{(2)}$ and enhanced QEC capability
of $\sab$. At $\gamma=0.1$, we compute the channel infidelities
for $\bin$ codes undergoing $\mathcal{N}_{\gamma}$ and optimal recovery
$\mathcal{R}^{\mathrm{o}}$ (obtained from convex optimization \cite{Fletcher2007,Albert2018})
and, in comparison, those for $\sab$ codes undergoing $\mathcal{R}^{(2)}$
and $\mathcal{R}^{\mathrm{o}}$, respectively, after $\mathcal{N}_{\gamma}$.
We note $\mathcal{R}^{\mathrm{o}}$ is considered here as it enables
the best code performance, yet it is only restricted to be completely
positive, trace preserving (CPTP) and can lack physics intuition and
ease of implementation.

We see from Fig.~\ref{fig:optimal infidelities for bin and sab}(a)
that, at $\gamma=0.1$, desired $\bin$ codes are found along $S\approx2N$
while the entire $N>S$ region features poor error protection. In
comparison, as manifested in Fig.~\ref{fig:optimal infidelities for bin and sab}(b),
$\sab$ codes with $\mathcal{R}^{(2)}$ achieve much lower infidelities
overall and open up the $N>S$ region where destructive interference
is pronounced. Fig.~\ref{fig:optimal infidelities for bin and sab}(c)
shows the minimum channel infidelities for $\sab$ codes under $\mathcal{R}^{\mathrm{o}}$.
Comparing Fig.~\ref{fig:optimal infidelities for bin and sab}(b)
and (c), we see that $\mathcal{R}^{(2)}$ suffices to unleash the
potential of $\sab$ at small $\gamma$, yielding infidelities close
to optimal. The results also demystify $\mathcal{R}^{\mathrm{o}}$
for single-mode codes with parity -- it is critical to capture the
orthogonal component between partially overlapping error words. An
alternative way to understand how SA helps suppress decoherence is
to decompose effective QEC-protected qubit channel $\mathcal{E}$
into Kraus operators \cite{SupplementalMaterial}. As shown in Fig.~\ref{fig:optimal infidelities for bin and sab}(d-f),
the advantage of $\sab$ code becomes more significant at $\gamma=0.25$,
with the emergence of a new desired regime instead of $S\approx2N$.
Since $\mathcal{R}^{(2)}$ only provides two layers of correction,
it begins to perform sub-optimally in regions where $\bar{n}\gamma\apprge S$
(or $N\gamma\apprge2$) at higher loss rate {[}Fig.~\ref{fig:optimal infidelities for bin and sab}(e){]},
indicating the need to resolve higher order error codewords or even
consider the optimal recovery. Nonetheless, $\sab$ codes with $\mathcal{R}^{(2)}$
still outperform $\bin$ codes with $\mathcal{R}^{\mathrm{o}}$, offering
a practically favorable and feasible QEC scheme.

$\cat$ code shares the parity structure with $\bin$ code and can
be similarly enhanced (detailed in \cite{SupplementalMaterial}).
Specifically, two-component $\cat$ code \cite{Mirrahimi2014}, the
simplest of the family which does not correct any excitation loss,
will correct one loss approximately after SA is imposed. Same as $\sab$,
the SA can be implemented by $\hat{U}_{\mathrm{S}}$ while now $2S$
is the number of coherent states in superposition -- this explains
why a small amount of cavity Kerr improves $\cat$'s performance (\cite{Albert2018},
Fig.~9a).

\paragraph*{Optimality vs.~complex noise}

As it is not obvious that periodic SA is the \textit{optimal} modification
for $\bin$, we relax the Kerr $\hat{U}_{\mathrm{K}}$ added to the
encoding to be general (detailed in Sec.~3 of \cite{SupplementalMaterial}).
At large $\gamma$ and $N\gg S$ regimes, indeed we are able to find
modified $\bin$ codes that further suppress the infidelities. These
results point to the potential benefit of going beyond RCQC to consider
generic phases in codeword designs {[}Fig.~\ref{fig:Graphical illustration of destructive interference}(c){]}.

To find the optimal encoding schemes given an error process, we deploy
the technique energy-constrained \cite{Noh2018} biconvex optimization
of recovery and encoding for channel fidelity \cite{Reimpell2005,Kosut2009}.
Given the encoding and error channel $\mathcal{N}\circ\mathcal{S}$,
computing optimal recovery $\mathcal{S}^{-1}\circ\mathcal{R}$ to
maximize channel fidelity is a convex problem, and vice versa. Using
this technique, we can compute the optimal RCQCs and CCQCs given an
error channel, respectively, to see if nontrivial phases are necessary
in optimal code constructs \cite{globalopt}.

For excitation loss, we find that optimal RCQCs and CCQCs are equivalent
up to global rotations, indicating the redundancy of nontrivial phases
\cite{SupplementalMaterial}. The observation that neither channel
imposes complex phases, reflected by their Kraus operators, confirms
that the nature of error determines the role of phase in optimal code
construction. 
\begin{table}[tb]
\centering{}%
\begin{tabular}{|c|>{\centering}p{1.6cm}|>{\centering}p{1.6cm}|>{\centering}p{1.6cm}|>{\centering}p{1.6cm}|}
\hline 
$Kt$ &
0 &
0.5 &
1 &
1.5\tabularnewline
\hline 
RCQCs &
\includegraphics[scale=0.15]{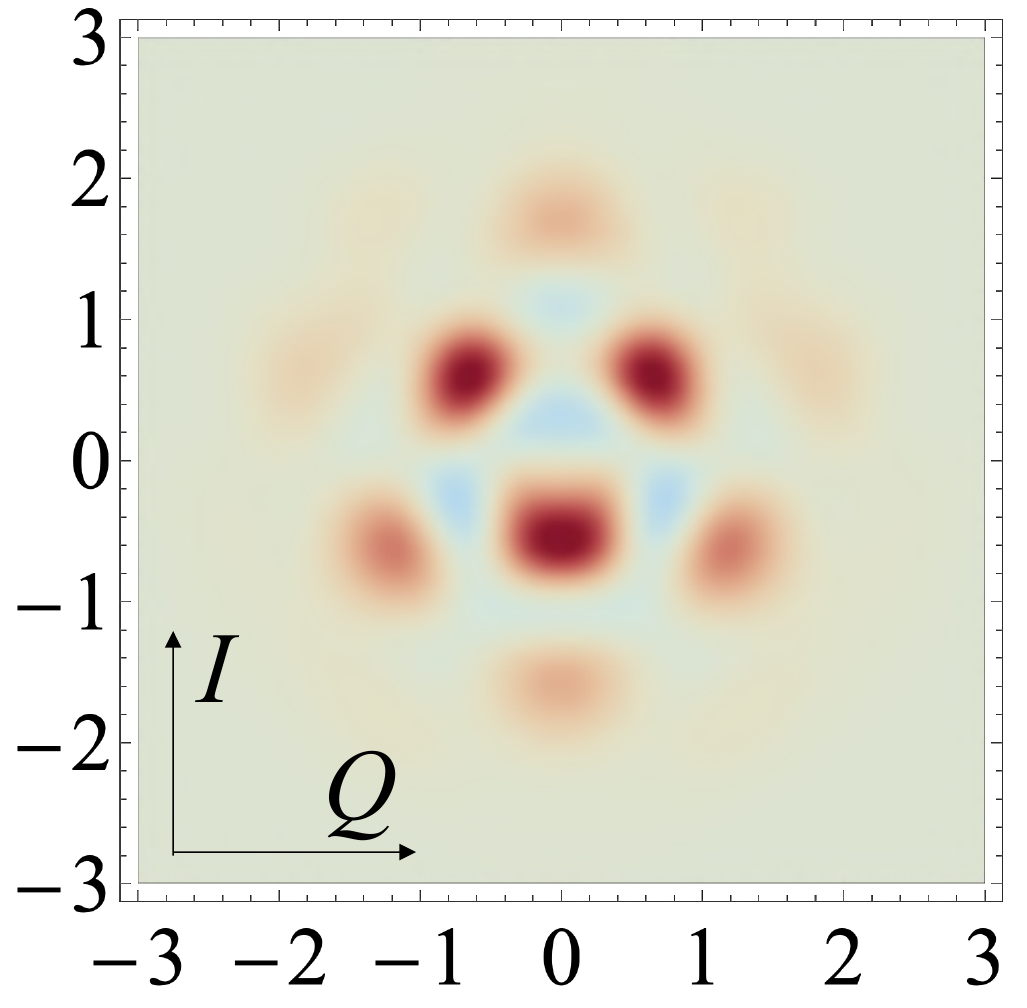}

6.1e-3 &
\includegraphics[scale=0.15]{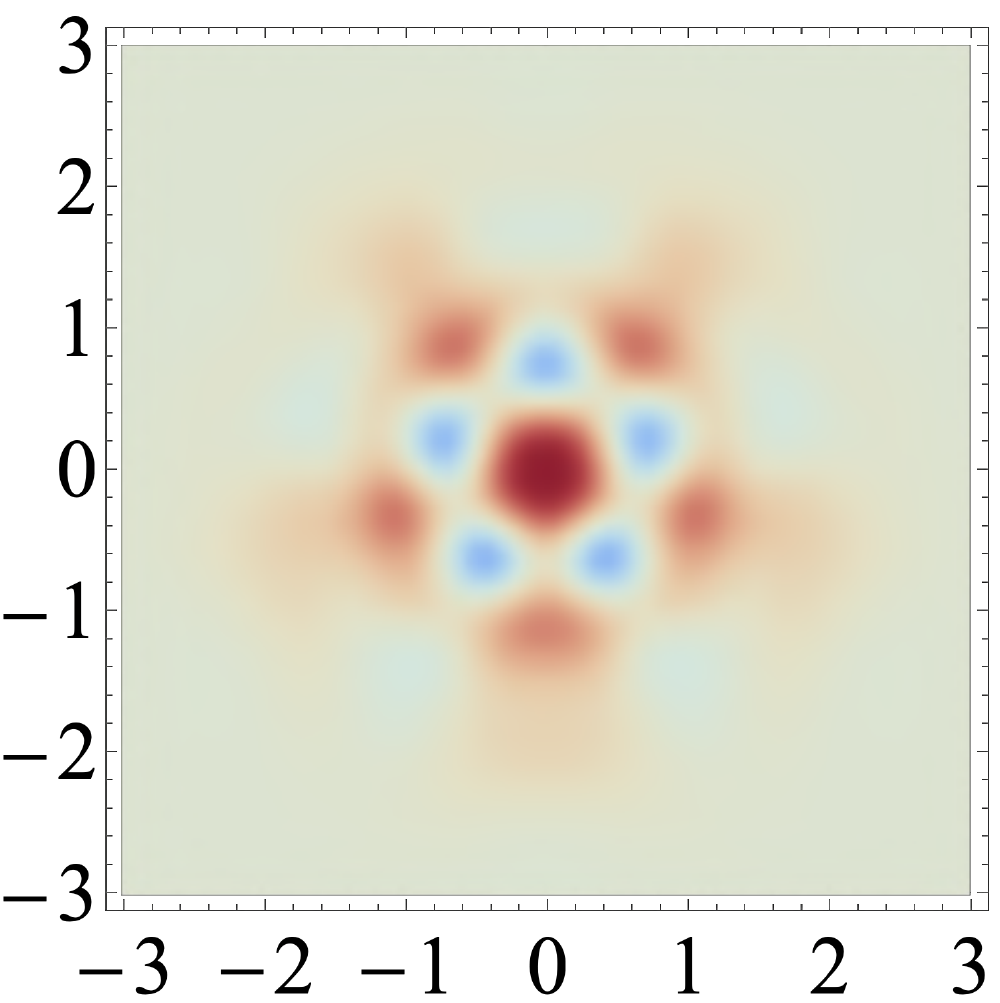}

1.4e-2 &
\includegraphics[scale=0.12]{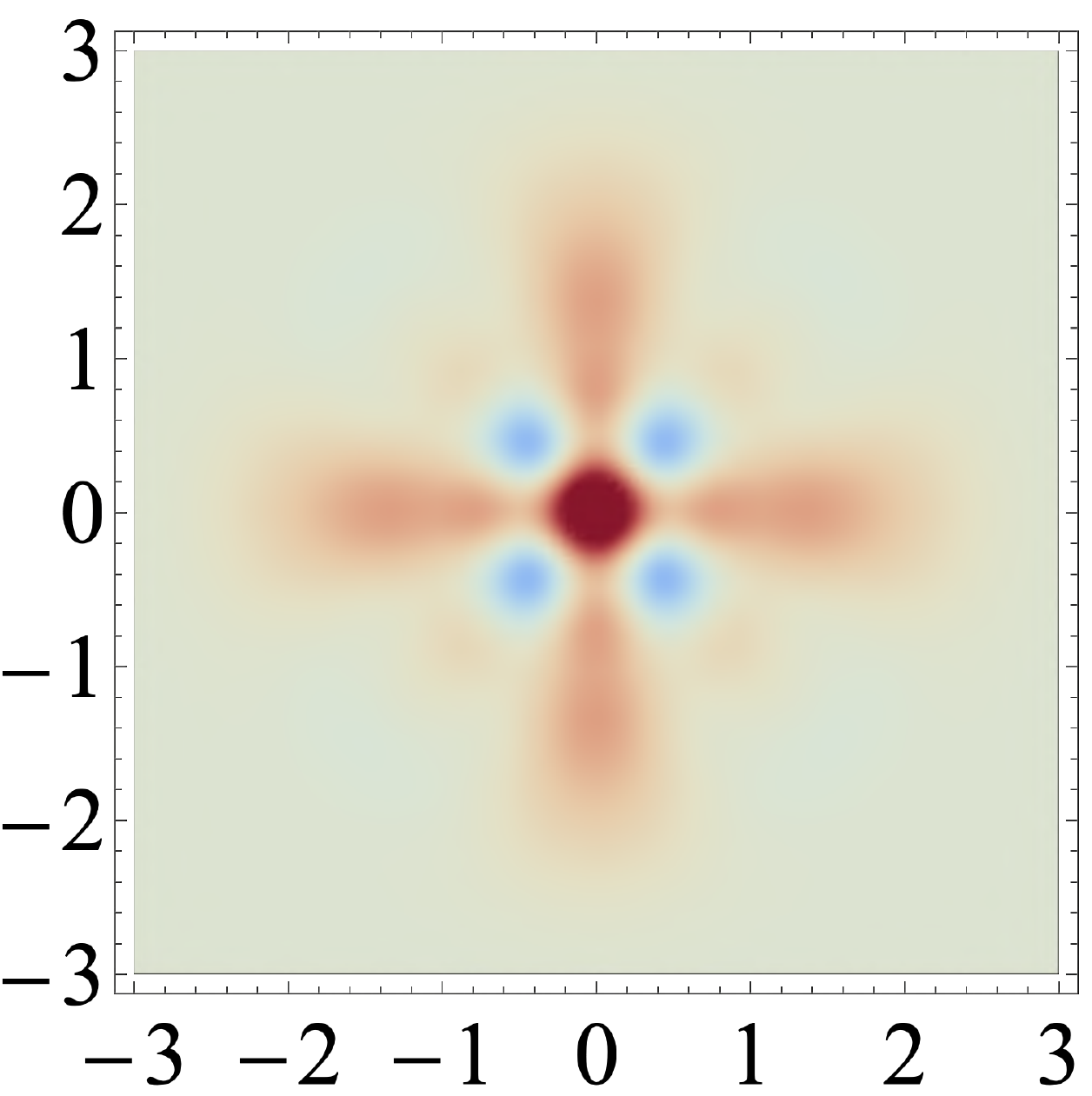}

2.5e-2 &
\includegraphics[scale=0.15]{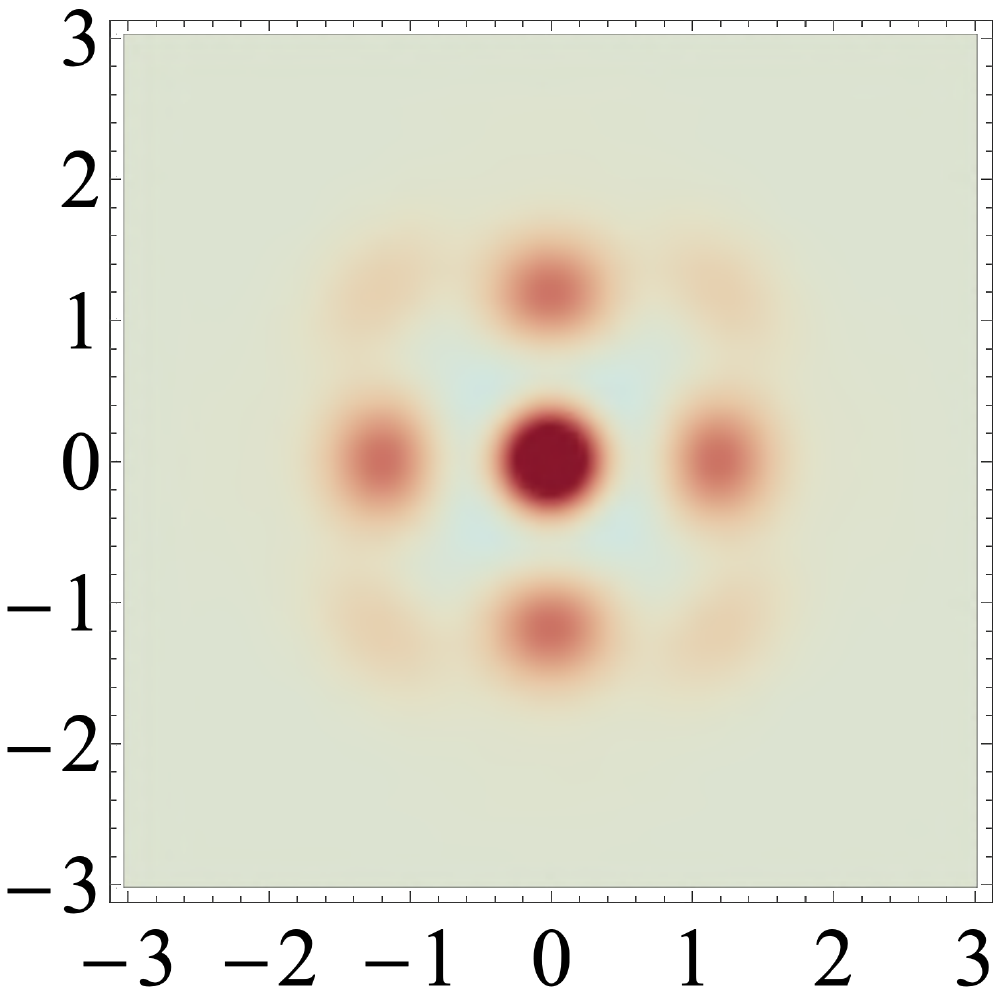}

3.6e-2\tabularnewline
\hline 
CCQCs &
\includegraphics[scale=0.12]{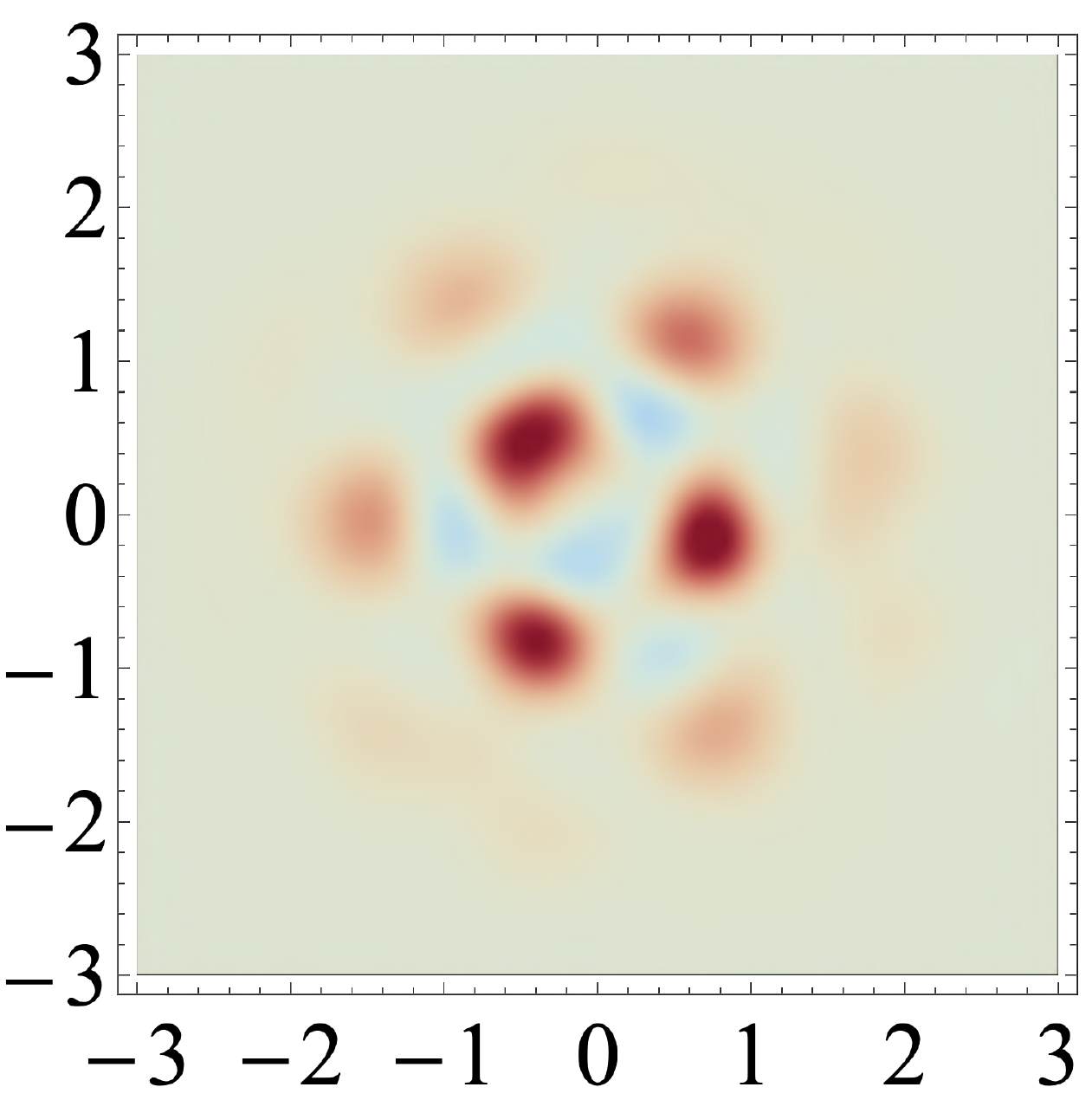}

6.1e-3 &
\includegraphics[scale=0.12]{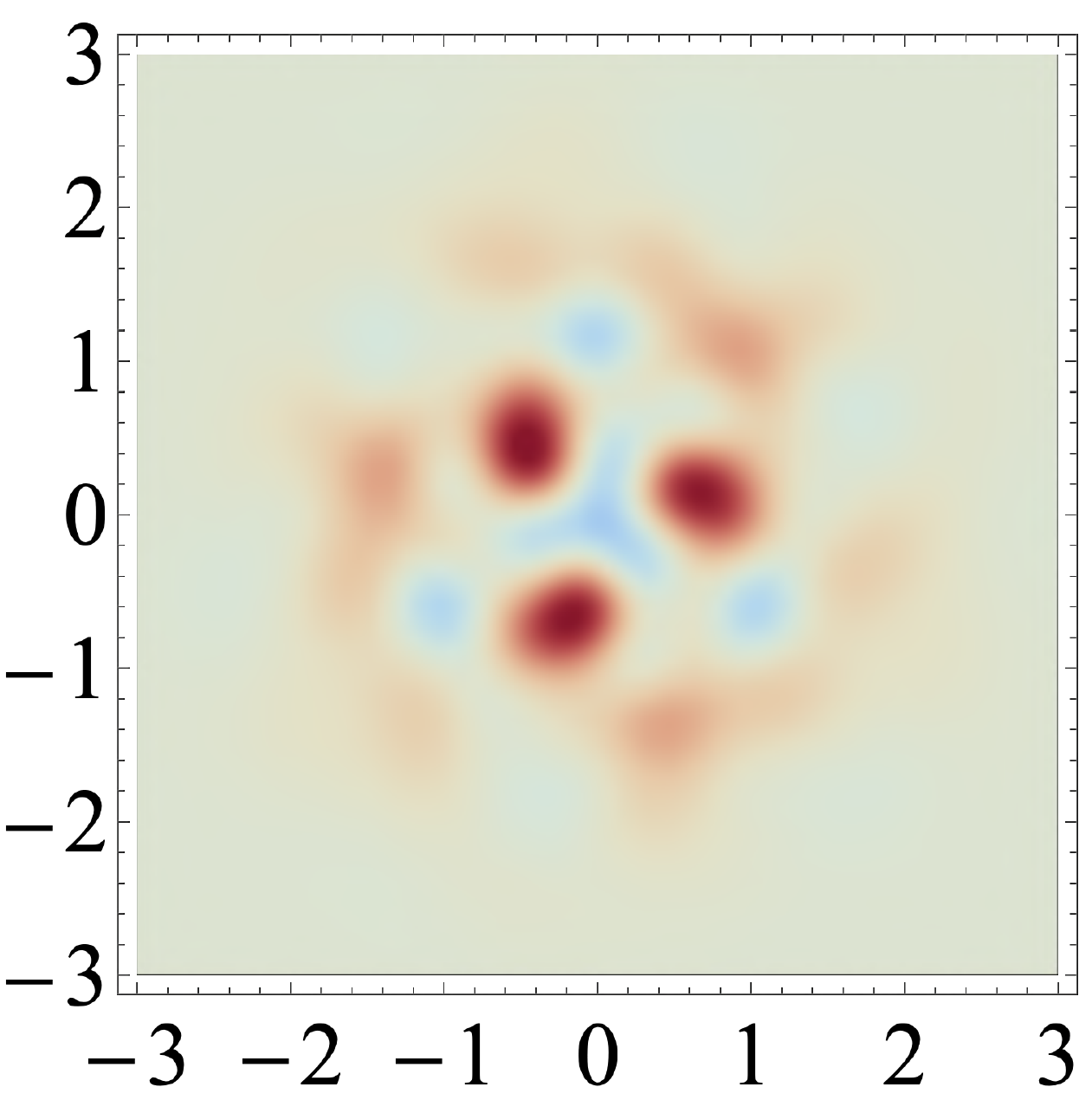}

1.2e-2 &
\includegraphics[scale=0.12]{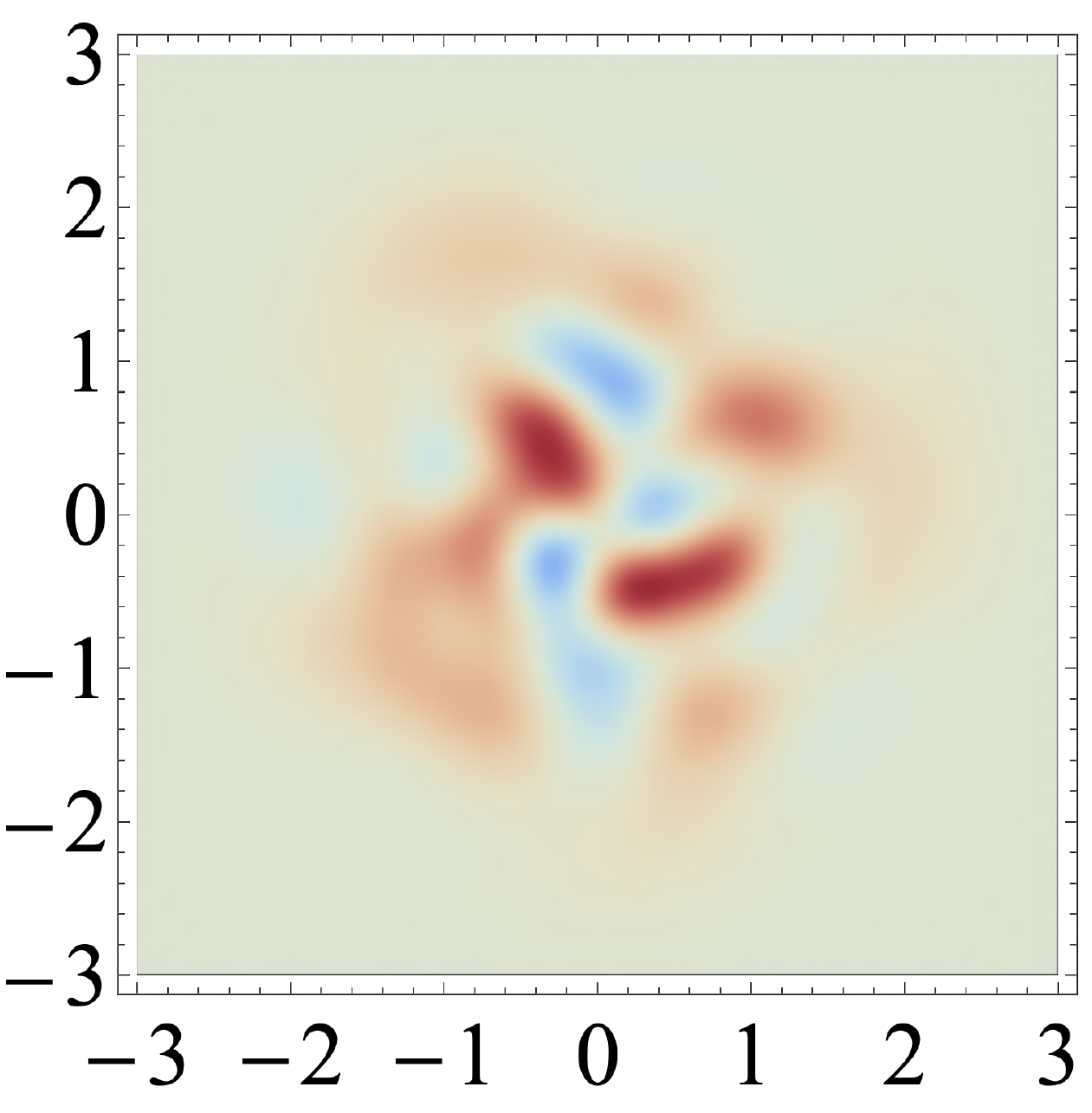}

2.3e-2 &
\includegraphics[scale=0.15]{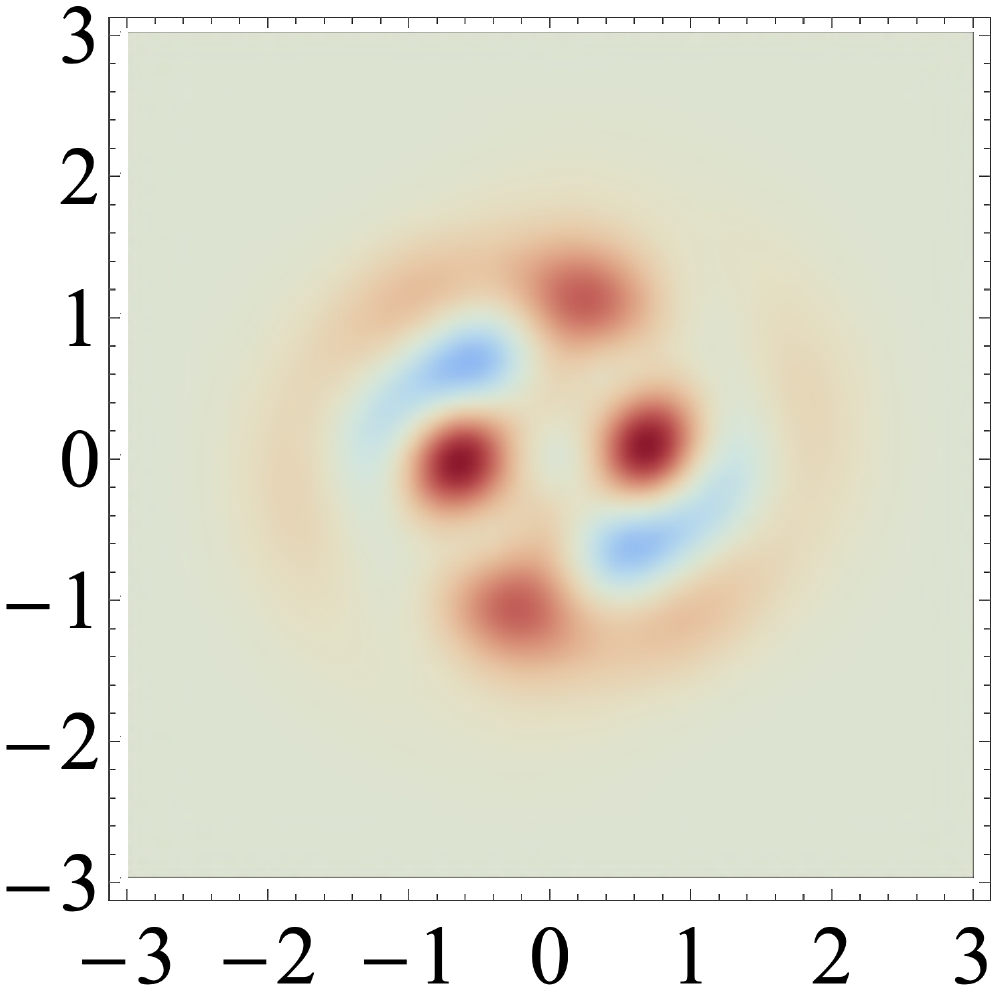}

3.2e-2\tabularnewline
\hline 
$\bin(2,2)$ &
1.8e-2 &
2.1e-2 &
3.0e-2 &
4.5e-2\tabularnewline
\hline 
\end{tabular}\caption{Results from biconvex optimization under energy constraint $\bar{n}_{\mathrm{c}}=2$
for single-mode RCQCs and CCQCs at $\gamma=0.1$ and different $Kt$.
Wigner functions of maximally mixed state $\frac{1}{2}\hat{P}_{\protect\cc}$
are shown for the obtained optimal codes, respectively, with associated
channel infidelities $1-F^{\mathrm{o}}$ below them. \label{tab:biconvex optimization results }The
infidelities for $\protect\bin(2,2)$ are shown for comparison.}
\end{table}
To see where nontrivial phases play a role, we consider the joint
channel of excitation loss and Kerr effect. The channel is relevant
(e.g. in circuit QED systems) due to the concurrence of cavity Kerr
and intrinsic loss. Kerr alone is reversible, but as it does not commute
with $\mathcal{N}_{\gamma}$, randomness in the timing of excitation
loss leads to uncertainty in phase $\hat{U}_{\mathrm{Kr}}$ imprints
and thus decoherence \cite{Hu2018}. The channel's superoperator \cite{Kerr}
is 
\begin{equation}
\mathcal{N}_{\gamma,\,Kt}=e^{-\frac{1}{2}iKt[\hat{n}^{2},\,\cdot]-\ln(1-\gamma)(\hat{a}\,\cdot\,\hat{a}^{\dagger}-\frac{1}{2}\{\hat{n},\,\cdot\})}\,.
\end{equation}
In Table~\ref{tab:biconvex optimization results } we show the Wigner
functions for the maximally mixed state $\frac{1}{2}\hat{P}_{\cc}$,
which contains full information of a code, and the associated infidelities
$1-F^{\mathrm{o}}$ of the optimized RCQCs and CCQCs, respectively,
at $\gamma=0.1$, $\bar{n}_{\mathrm{c}}=2$ and various $Kt$. As
a comparison we show the infidelities for $\bin(2,2)$ with $\bar{n}=2$,
which protects against pure excitation loss \cite{Albert2018}. For
all $Kt$, both scenarios yield codes better than $\bin(2,2)$, indicating
the advantage of codes tailored to $\mathcal{N}_{\gamma,\,Kt}$. At
$Kt=0$ (i.e. pure excitation loss) the optimized RCQC and CCQC are
equivalent, as discussed earlier. For $Kt\neq0$, nontrivial phases
allow optimal CCQCs to offer better protection. The results exemplify
the necessity of CQCCs for practical optimal QEC, as phase-imposing
errors can often co-exist with other decoherence sources that they
do not commute with.

\paragraph*{Sign-altered $\protect\gkp$ and multi-qubit codes}

So far, we have focused on codes defined in the Fock basis, while
the same idea of phase engineering should apply to code constructs
in other bases. For example, one can modify $\gkp$ code, defined
in phase space for drift error \cite{Gottesman2001}, as follows 
\begin{equation}
|\mu_{\sigma}\ket=\underset{s\in\mathbb{Z}}{\sum}(-1)^{\sigma}|q=\alpha(\sigma+2s)\ket
\end{equation}
where $\sigma=0,1$ and $2\alpha$ is the spacing between position
eigenstates. Similar to $\sab$ and $\sac$, the SA can be imposed
by following the original $\gkp$ encoding procedure with $e^{i\pi\hat{q}^{2}/4\alpha^{2}}$,
which transforms the $\gkp$ stabilizer $\hat{S}_{1}=e^{-ip\alpha}$
to $\hat{S}'_{1}=e^{-i(p-\pi q/2\alpha^{2})\alpha}$ while leaving
$\hat{S}_{2}=e^{2\pi iq/\alpha}$ unchanged. The new stabilizers define
a nonrectangular $\gkp$ lattice, and for $\alpha^{2}=\sqrt{3}\pi/2$
it is hexagonal -- the optimal packing in two dimensions with a larger-in-size
smallest uncorrectable shift \cite{Gottesman2001,Harrington2001}.

The interference effects further extend into the multi-qubit regime.
Consider Shor's $[[9,1,3]]$ code that corrects arbitrary single-qubit
Pauli errors with codewords $|-_{\sh}\ket\propto|\tilde{1}\tilde{0}\tilde{0}\ket+|\tilde{0}\tilde{1}\tilde{0}\ket+|\tilde{0}\tilde{0}\tilde{1}\ket+|\tilde{1}\tilde{1}\tilde{1}\ket$
and $|+_{\sh}\ket=\s_{x}^{\otimes9}|-_{\sh}\ket$, where $\tilde{i}=iii$
stands for blocks of three qubits. It detects weight-three $\s_{x}$
errors, except for $\s_{x}^{(i)}\s_{x}^{(i+1)}\s_{x}^{(i+2)}$ with
$i=1,4,7$ as they are logical operators. We now consider a sign-altered
variant with $|-_{\shp}\ket\propto|\tilde{1}\tilde{0}\tilde{0}\ket-|\tilde{0}\tilde{1}\tilde{0}\ket+|\tilde{0}\tilde{0}\tilde{1}\ket-|\tilde{1}\tilde{1}\tilde{1}\ket$
and $|+_{\shp}\ket=|+_{\sh}\ket$. The new code detects $\s_{x}^{(i)}\s_{x}^{(i+1)}\s_{x}^{(i+2)}$
and thus all weight-three $\s_{x}$ errors. In addition, as detailed
in \cite{SupplementalMaterial}, it detects more weight-three hybrid
$\sigma_{x}$ and $\sigma_{y}$ errors while offers the same protection
over $\sigma_{z}$ as the original Shor code. A similar modification
can improve Shor and Steane codes over qubit amplitude damping, which
is a realistic concern for qubit systems \cite{SupplementalMaterial,Darmawan2017}.

\paragraph{Conclusion}

In contrast to conventional designs of quantum codes whose error-correction
capability comes from spanning codewords with distinct subsets of
computational basis states, we explored the conjugate degree of freedom,
the phases carried by basis states, to devise efficient quantum codes
for various bosonic and qubit errors. The new codes can feature destructive
interference and hence suppressed overlap between error codewords.
To showcase the principle, we modify the codewords of bosonic binomial
and cat codes that correct excitation loss errors by making certain
probability amplitudes negative. With a quantum recovery that effectively
captures the suppressed overlap, the modified codes demonstrate desired
error-correction performance. For complex-valued noises, such as the
joint channel of excitation loss and cavity Kerr, we show that considering
complex-valued amplitudes in codewords is critical for optimal code
constructs. The same principle also helps improve multi-qubit codes
in overcoming noises such as Pauli errors and qubit amplitude damping.
We expect the results developed here to deepen our understanding of
quantum error correction and enable development of efficient quantum
codes across a wide array of physical platforms for faithful quantum
information processing.

\paragraph{Acknowledgements}

We thank Philippe Faist, Mengzhen Zhang, Sisi Zhou and Yuhui Ouyang
for helpful discussions. We acknowledge supports from the ARL-CDQI
(W911NF-15-2-0067, W911NF-18-2-0237), ARO (W911NF-18-1-0020, W911NF-18-1-0212),
ARO MURI (W911NF-16-1-0349), AFOSR MURI (FA9550-14-1-0052, FA9550-15-1-0015),
DOE (DE-SC0019406), NSF (EFMA-1640959), and the Packard Foundation
(2013-39273).

\bibliographystyle{apsrev4-1}
\bibliography{13_Users_LLS_Dropbox_Research_Constructing_good___ructing_bosonic_codes_via_phase_engineering}

\begin{thebibliography}{39}
\bibitem [{glo()}]{globalopt}%
  \BibitemOpen
  \href@noop {} {}\bibinfo {note} {We note that biconvex optimization does not
  guarantee globally optimality. The optimal codes are found via sampling
  across initial conditions}\BibitemShut {NoStop}%
\end{thebibliography}

\end{document}


\global\long\def\half{\frac{1}{2}}%
\global\long\def\a{\alpha}%
\global\long\def\b{\beta}%
\global\long\def\g{\gamma}%
\global\long\def\c{\chi}%
\global\long\def\d{\delta}%
\global\long\def\o{\omega}%
\global\long\def\m{\mu}%
\global\long\def\s{\sigma}%
\global\long\def\n{\nu}%
\global\long\def\z{\zeta}%
\global\long\def\l{\lambda}%
\global\long\def\k{\kappa}%
\global\long\def\x{\chi}%
\global\long\def\r{\rho}%
\global\long\def\t{\theta}%
\global\long\def\G{\Gamma}%
\global\long\def\D{\Delta}%
\global\long\def\O{\Omega}%
\global\long\def\pr{\prime}%
\global\long\def\k{\kappa}%
\global\long\def\S{\mathcal{S}}%
\global\long\def\R{\mathcal{R}}%
\global\long\def\E{\mathcal{E}}%
\global\long\def\bra{\langle}%
\global\long\def\ket{\rangle}%
\global\long\def\dg{\dagger}%
\global\long\def\dgt{\ddagger}%
\global\long\def\tr{\text{Tr}}%
\global\long\def\Tr{\textsc{Tr}}%
\global\long\def\id{\mathcal{I}}%
\global\long\def\e{\bar{\eta}}%
\global\long\def\nb{\bar{n}}%
\global\long\def\ph{\hat{n}}%
\global\long\def\aa{\hat{a}}%
\global\long\def\pc{P_{\!\textnormal{\ensuremath{\mathtt{cat}}}}}%
\global\long\def\pl{P_{\!\textnormal{\ensuremath{\mathtt{gkp2}}}}}%
\global\long\def\pb{P_{\textnormal{\ensuremath{\mathtt{bin}}}}}%
\global\long\def\pg{P_{\!\textnormal{\ensuremath{\mathtt{gkp}}}}}%
\global\long\def\pn{P_{\!\textnormal{\ensuremath{\mathtt{num}}}}}%
\global\long\def\pp{P_{\!\textnormal{\ensuremath{\mathtt{code}}}}}%
\global\long\def\gkp{\textnormal{\ensuremath{\mathtt{gkp}}}}%
\global\long\def\cat{\textnormal{\ensuremath{\mathtt{cat}}}}%
\global\long\def\cc{\textnormal{\ensuremath{\mathtt{code}}}}%
\global\long\def\num{\textnormal{\ensuremath{\mathtt{num}}}}%
\global\long\def\bin{\textnormal{\ensuremath{\mathtt{bin}}}}%
\global\long\def\lat{\textnormal{\ensuremath{\mathtt{gkp2}}}}%
\global\long\def\xx{\hat{x}}%
\global\long\def\fe{F_{e}}%
\global\long\def\A{\mathcal{A}}%
\global\long\def\L{\mathcal{L}}%
\global\long\def\sab{\textnormal{\ensuremath{\mathtt{sab}}}}%
\global\long\def\okb{\textnormal{\ensuremath{\mathtt{okb}}}}%
\global\long\def\sac{\textnormal{\ensuremath{\mathtt{sac}}}}%
\global\long\def\st{\textnormal{\ensuremath{\mathtt{stn^{\pr}}}}}%
\global\long\def\ot{\otimes}%
\global\long\def\er{\mathbb{E}}%
\global\long\def\sh{\textnormal{\ensuremath{\mathtt{shor}}}}%
\global\long\def\shp{\textnormal{\ensuremath{\mathtt{shor'}}}}%
\global\long\def\shpp{\textnormal{\ensuremath{\mathtt{shor''}}}}%
\global\long\def\lket#1{\left|#1\right\rangle }%

\title{Designing good bosonic quantum codes via creating destructive interference
-- supplementary material}
\maketitle

\section{Effective qubit channel decomposition}

In this section, we examine the four Kraus operators of optimal effective
qubit channels $\mathcal{E}_{\bin}^{\mathrm{o}}=\mathcal{S}_{\bin}^{-1}\circ\mathcal{R}_{\bin}^{\mathrm{o}}\circ\mathcal{N}_{\gamma}\circ\mathcal{S}_{\bin}$
and $\mathcal{E}_{\sab}^{\mathrm{o}}=\mathcal{S}_{\sab}^{-1}\circ\mathcal{R}_{\sab}^{\mathrm{o}}\circ\mathcal{N}_{\gamma}\circ\mathcal{S}_{\sab}$,
to visualize how SA leads to change in the leading error.

In Table~\ref{tab: Kraus operators of effective qubit channel},
we compare $\bin$ and $\sab$ with $N=4$, $S=3$ at $\gamma=0.1$.
We can see that the leading error for $\bin$ code with $\mathcal{R}^{\mathrm{o}}$
is a bit-flip error, resulted from events with more than $S-1$ losses.
On the other side, $\sab$ code with $\mathcal{R}^{\mathrm{o}}$ reduces
that bit-flip error by more than half -- The remainder is turned
into a $\sigma_{+}$-type error, as $|\bra1_{\mathrm{\sab}}|\hat{E}_{k}^{\dagger}\hat{E}_{S+k}|0_{\mathrm{\sab}}\ket|$
and $|\bra0_{\mathrm{\sab}}|\hat{E}_{k}^{\dagger}\hat{E}_{S+k}|1_{\mathrm{\sab}}\ket|$
in general differ due to the difference in basis state occupation
profiles between $|0_{\mathrm{\sab}}\ket$ and $|1_{\mathrm{\sab}}\ket$.
\begin{table}[H]
\centering{}%
\begin{tabular}{cccc}
\hline 
\noalign{\vskip0.1cm}
\multicolumn{2}{c}{$\mathcal{E}_{\bin}^{\mathrm{o}}$} &
\multicolumn{2}{c}{$\mathcal{E}_{\sab}^{\mathrm{o}}$}\tabularnewline
\hline 
\noalign{\vskip0.2cm}
Kraus operator &
Prob. &
Kraus operator &
Prob.\tabularnewline[0.2cm]
\hline 
\noalign{\vskip0.1cm}
$\left(\begin{array}{cc}
0.99 & 0\\
0 & 0.99
\end{array}\right)$ &
9.8e-1 &
$\left(\begin{array}{cc}
1.0 & 0\\
0 & 0.99
\end{array}\right)$ &
9.9e-1\tabularnewline[0.1cm]
\noalign{\vskip0.1cm}
$\left(\boldsymbol{\begin{array}{cc}
0 & -0.15\\
-0.16 & 0
\end{array}}\right)$ &
\textbf{2.4e-2} &
$\left(\boldsymbol{\begin{array}{cc}
0 & -0.13\\
-0.0098 & 0
\end{array}}\right)$ &
\textbf{8.2e-3}\tabularnewline[0.1cm]
\noalign{\vskip0.1cm}
$\left(\begin{array}{cc}
-0.023 & 0\\
0 & 0.023
\end{array}\right)$ &
5.3e-4 &
$\left(\begin{array}{cc}
0 & -0.0031\\
0.040 & 0
\end{array}\right)$ &
8.0e-4\tabularnewline[0.1cm]
\noalign{\vskip0.1cm}
$\left(\begin{array}{cc}
0 & -0.0043\\
0.0040 & 0
\end{array}\right)$ &
1.7e-5 &
$\left(\begin{array}{cc}
-0.028 & 0\\
0 & 0.028
\end{array}\right)$ &
7.9e-4\tabularnewline[0.1cm]
\hline 
\end{tabular}\caption{Kraus operators (arranged in the descending order of probabilities)
of the optimal effective qubit channel for $\protect\bin(4,3)$ and
$\protect\sab(4,3)$ at $\gamma=0.1$.\label{tab: Kraus operators of effective qubit channel}
The key difference between the two qubit channels, i.e. the suppression
of bit-flip error, is highlighted in bold.}
\end{table}

\section{Further examples of sign alteration}

\subsection{Modified cat code for excitation loss\label{subsec:Sign-altered cat code}}

Consider the simplest $\cat$ code $|\s_{\cat}\ket\propto\sum_{n=0}^{\infty}\frac{\a^{2n+\s}}{\sqrt{(2n+\s)!}}|2n+\s\ket$
for $\s\in\{0,1\}$. This PCQC cannot detect even a single loss event
$\aa$, as $\aa|1_{\cat}\ket\propto|0_{\cat}\ket$ and visa versa.
However, a simple modification produces the sign-altered cat ($\sac$)
codewords $|\s_{\sac}\ket\propto\sum_{n=0}^{\infty}\frac{\a^{2n+\s}}{\sqrt{(2n+\s)!}}(-1)^{\s+1}|2n+\s\ket$
that do not significantly overlap upon a loss event. While $\aa|1_{\sac}\ket$
is still in the support of $|0_{\sac}\ket$ (i.e., in the even Fock
state subspace), the overlap between $\aa|1_{\sac}\ket$ and $|0_{\sac}\ket$
is exponentially suppressed with $\a$ due to \textit{destructive
interference}. Thus, SA can extend a $\cat$ code that does not correct
a loss error to one that approximately does. This behavior can also
be understood in terms of bosonic coherent states $|\a\ket$ (with
$\aa|\a\ket=\a|\a\ket$). In this basis, $|0_{\sac}\ket\propto|i\a\ket+\lket{-i\a}$,
$|1_{\sac}\ket\propto|\a\ket-\lket{-\a}$, and $\bra0_{\sac}|\aa|1_{\sac}\ket\rightarrow0$
since the coherent states of $|0_{\sac}\ket$ are well separated from
those of $\aa|1_{\sac}\ket$.

Generically, the codewords of $\cat$ codes are cat states, superpositions
of $2S$ coherent states lying equidistantly on a circle in the phase
space \citet{Li2017}. In Fock basis, they can be expressed as
\begin{eqnarray}
|C_{\alpha}^{n}\ket & = & \sqrt{\frac{2S}{\mathcal{N}_{\alpha}^{n}}}\stackrel[m=0]{\infty}{\sum}\frac{e^{-\frac{\alpha^{2}}{2}}\alpha^{n+2mS}}{\sqrt{(n+2mS)!}}|n+2mS\ket
\end{eqnarray}
where $n=0,1,\ldots,2S-1$, $\mathcal{N}_{\alpha}^{n}=2S\bra\alpha|\hat{\Pi}_{n\bmod S}|\alpha\ket$
is the normalization factor, and $\underset{\alpha\rightarrow\infty}{\lim}\mathcal{N}_{\alpha}^{n}=1$.
Without losing generality, $\alpha$ is assumed to be real here and
we choose the $\mathrm{Span}\{|C_{\alpha}^{0}\ket,|C_{\alpha}^{S}\ket\}$
to be the logical subspace -- for $\cat$ codes there is freedom
on the choice of ``$d$-subspace'', $\mathrm{Span}\{|C_{\alpha}^{d}\ket,|C_{\alpha}^{d+S}\ket\}$
as the logical subspace (\citet{Li2017}).

Generic $\sac$ code is defined as follows
\begin{eqnarray}
|0_{\sac}\ket & = & \sqrt{\frac{2S}{\mathcal{N}_{\alpha}^{0}}}\stackrel[m=0]{\infty}{\sum}(-1)^{m}\frac{e^{-\frac{\alpha^{2}}{2}}\alpha^{2mS}}{\sqrt{2mS!}}|2mS\ket\,,\label{eq:sac 0-logical}\\
|1_{\sac}\ket & = & \sqrt{\frac{2S}{\mathcal{N}_{\alpha}^{S}}}\stackrel[m=0]{\infty}{\sum}\frac{e^{-\frac{\alpha^{2}}{2}}\alpha^{(2m+1)S}}{\sqrt{(2m+1)S!}}|(2m+1)S\ket\,.\label{eq:sac 1-logical}
\end{eqnarray}

\begin{figure}[tb]
\centering{}\includegraphics[scale=0.5]{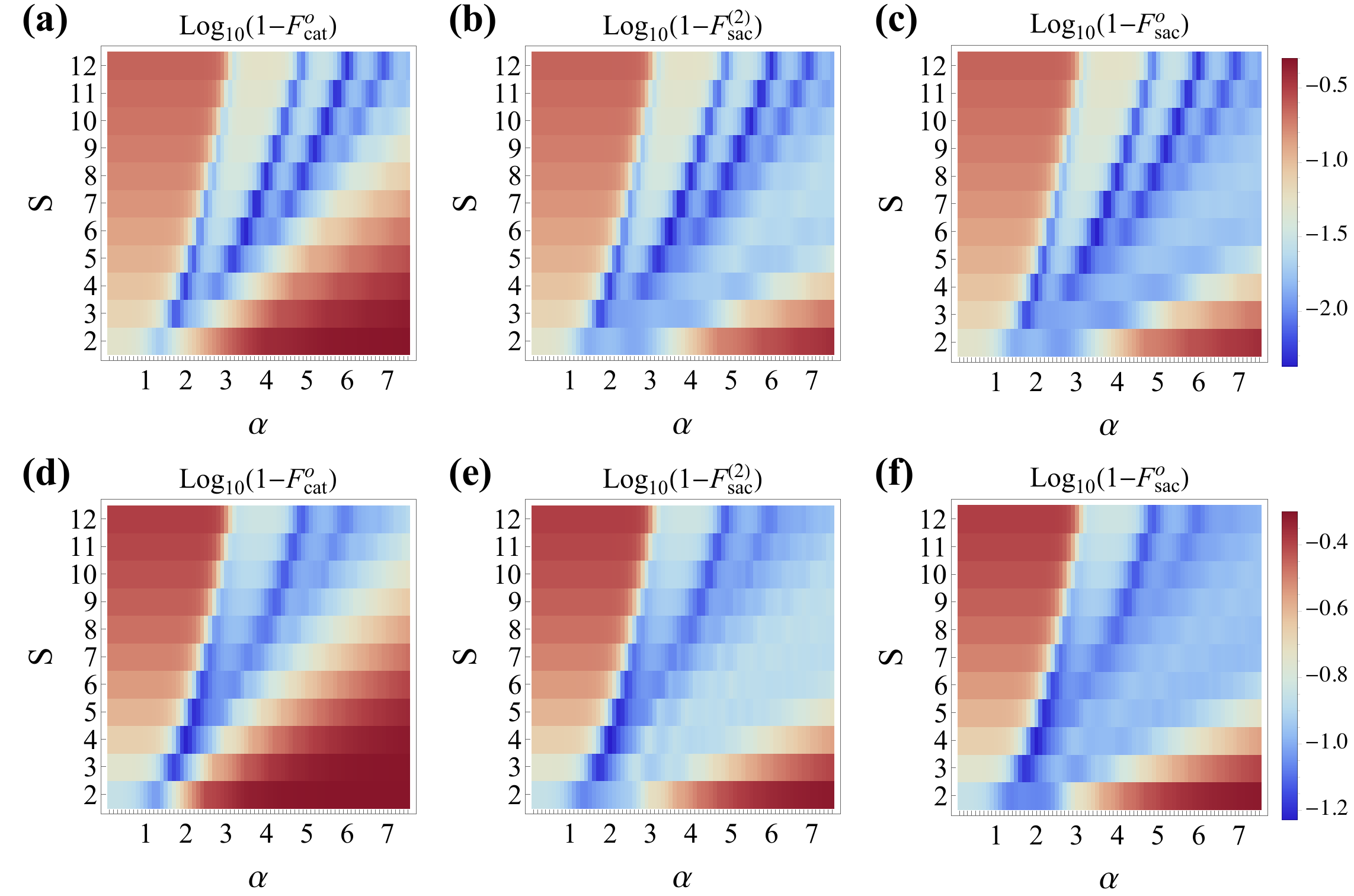}\caption{Channel infidelities (in logarithmic scale) for (a-c) $\protect\cat$
code with optimal recovery (as indicated by the superscript), $\protect\sac$
code with two-level recovery and $\protect\sac$ code with optimal
recovery, respectively, at $\gamma=0.1$, and (d-f) same as (a-c)
except for $\gamma=0.25$. Each point represents a code with associated
$S$ and $\alpha$. \label{fig:Optimally infidelity-1}}
\end{figure}
The above codewords can be generated in the same way as $\sab$ code,
i.e. $\mathcal{S}_{\sac}(\cdot)=\hat{U}_{\mathrm{S}}\mathcal{S}_{\cat}(\cdot)\hat{U}_{\mathrm{S}}^{\dagger}$.
We can see from Fig.~\ref{fig:Optimally infidelity-1} that for $\sac$
the $\alpha^{2}\gamma\approx S$ region is opened up due to the additional
suppression of errors induced by $S$ to $2S-1$ losses. For small
loss, $\mathcal{R}^{(2)}$ yields QEC performance close to optimal.
Unlike $\bin$ codes, at finite $\alpha$, the performance of $\cat$
codes is also constrained by dephasing caused by $\bra0_{\sac}|\hat{n}^{m}|0_{\sac}\ket\neq\bra1_{\sac}|\hat{n}^{m}|1_{\sac}\ket$
for $m\in\mathbb{Z}^{+}$. This is why at low loss rates the optimal
codes are found around ``sweet spots'' that are specific choices
of $\alpha$ minimizing the dephasing (\citet{Li2017}). As a result,
SA that suppresses bit-flip type of decoherence cannot bring improvement
in QEC performance as significant as it does for $\bin$ codes.

\subsection{Modified Shor's $[[9,1,3]]$ code for Pauli errors}

Given a code with projection $P$ and error $E$, the error is said
to be \textit{detectable} if $PEP\propto P$. A set of errors $\er=\{E\}$
is said to \textit{correctable} if $PE^{\dg}FP\propto P$ for all
$E,F\in\er$. Here we showcase a sign-altered variant of Shor code
that better protects against Pauli errors.

With stabilizers $\sigma_{z}^{(1)}\sigma_{z}^{(2)}$, $\sigma_{z}^{(2)}\sigma_{z}^{(3)}$,
$\sigma_{z}^{(4)}\sigma_{z}^{(5)}$, $\sigma_{z}^{(5)}\sigma_{z}^{(6)}$,
$\sigma_{z}^{(7)}\sigma_{z}^{(8)}$, $\sigma_{z}^{(8)}\sigma_{z}^{(9)}$,
$\s_{x}^{(1)}\s_{x}^{(2)}\s_{x}^{(3)}\s_{x}^{(4)}\s_{x}^{(5)}\s_{x}^{(6)}$
and $\s_{x}^{(4)}\s_{x}^{(5)}\s_{x}^{(6)}\s_{x}^{(7)}\s_{x}^{(8)}\s_{x}^{(9)}$,
Shor code
\begin{equation}
\begin{cases}
|+_{\sh}\ket & =\frac{1}{2}(|000\,000\,000\ket+|000\,111\,111\ket+|111\,000\,111\ket+|111\,111\,000\ket)\\
|-_{\sh}\ket & =\frac{1}{2}(|111\,000\,000\ket+|000\,111\,000\ket+|000\,000\,111\ket+|111\,111\,111\ket)
\end{cases}\label{eq:shor code}
\end{equation}
corrects arbitrary single-qubit errors. For two-qubit errors, the
code only offers partial correction as $\s_{x}^{(i)}\s_{x}^{(i+1)}\s_{x}^{(i+2)}$
with $i=1,4,7$ are logical $\sigma_{x}$ operators. We then consider
the following modified Shor code
\begin{equation}
\begin{cases}
|+_{\shp}\ket & =\frac{1}{2}(|000\,000\,000\ket-|000\,111\,111\ket+|111\,000\,111\ket-|111\,111\,000\ket)\\
|-_{\shp}\ket & =\frac{1}{2}(|111\,000\,000\ket+|000\,111\,000\ket+|000\,000\,111\ket+|111\,111\,111\ket)
\end{cases}\,.
\end{equation}
The stabilizer set of this code consists of the same six $\sigma_{z}$
stabilizers as Shor code, $\s_{x}^{(1)}\s_{x}^{(2)}\s_{x}^{(3)}\s_{x}^{(7)}\s_{x}^{(8)}\s_{x}^{(9)}$
and $\s_{y}^{(1)}\s_{x}^{(2)}\s_{x}^{(3)}\s_{y}^{(4)}\s_{x}^{(5)}\s_{x}^{(6)}\sigma_{z}^{(7)}$.
Its logical operators are $\bar{\sigma}_{x}=\s_{z}^{\otimes9}$ and
$\bar{\sigma}_{z}=\s_{x}^{(1)}\s_{x}^{(2)}\s_{x}^{(3)}\s_{z}^{(4)}$,
which is now weight-four. Same as Shor code, the modified code corrects
arbitrary single-qubit and detects arbitrary two-qubit Pauli errors.
In addition, one can see that
\begin{equation}
\begin{cases}
\s_{x}^{(1)}\s_{x}^{(2)}\s_{x}^{(3)}|+_{\shp}\ket & =\frac{1}{2}(|111\,000\,000\ket-|000\,111\,000\ket+|000\,000\,111\ket-|111\,111\,111\ket)\\
\s_{x}^{(1)}\s_{x}^{(2)}\s_{x}^{(3)}|-_{\shp}\ket & =\frac{1}{2}(|000\,000\,000\ket+|000\,111\,111\ket+|111\,000\,111\ket+|111\,111\,000\ket)
\end{cases}\,,
\end{equation}
which leads to $P\s_{x}^{(1)}\s_{x}^{(2)}\s_{x}^{(3)}P=0$. In fact,
the code now detects $E=\s_{x}^{(i)}\s_{x}^{(i+1)}\s_{x}^{(i+2)}$
for $i=1,4,7$ and hence all weight-three $\sigma_{x}$ errors. Similarly,
one can check that fewer weight-three hybrid $\sigma_{x}$ and $\sigma_{y}$
errors are now undetectable. For error correction, one can also design
recoveries capture the additional QEC matrix elements that are now
zero. The trade-off for the improved error detection/correction is
that the new code is no longer a CSS code.

\subsection{Modified Shor's $[[9,1,3]]$ code and Steane's $[[7,1,3]]$ code
for qubit amplitude damping}

For single-qubit amplitude damping channel with rate $\g$, the Kraus
operators are 
\begin{align}
A_{0} & =I+\left(\sqrt{1-\g}-1\right)\s_{+}\s_{-}\,\,\,\,\,\,\,\,\,\,\,\,\,\,\,\,\,\,\,\,\,\,\,\,\text{and}\,\,\,\,\,\,\,\,\,\,\,\,\,\,\,\,\,\,\,\,\,\,\,\,A_{1}=\sqrt{\g}\s_{-}
\end{align}
where $\s_{-}=|0\ket\bra1|=\s_{+}^{\dg}$.

Consider correction of qubit amplitude damping errors with Shor code
in Eq.~\ref{eq:shor code}. The logical codewords can be conveniently
expressed as $|-\ket\propto|\tilde{1}\tilde{0}\tilde{0}\ket+|\tilde{0}\tilde{1}\tilde{0}\ket+|\tilde{0}\tilde{0}\tilde{1}\ket+|\tilde{1}\tilde{1}\tilde{1}\ket$
and $|+_{\sh}\ket=\s_{x}^{\otimes9}|-_{\sh}\ket$ where $\tilde{i}=iii$
stands for blocks of three qubits. The code detects arbitrary two-qubit
damping errors and ceases to protect some of the three-qubit errors,
such as $\tilde{\s}_{-}^{(i)}=\s_{-}^{(i)}\s_{-}^{(i+1)}\s_{-}^{(i+2)}$
for $i\in\{1,4,7\}$ and $\s_{-}^{(i)}=|0\ket_{i}\bra1|_{i}$. Now
consider the modified codeword $|-_{\shpp}\ket\propto|\tilde{1}\tilde{0}\tilde{0}\ket+|\tilde{0}\tilde{1}\tilde{0}\ket+|\tilde{0}\tilde{0}\tilde{1}\ket-|\tilde{1}\tilde{1}\tilde{1}\ket$.
Taking $i=1$ as an example, $\tilde{\s}_{-}^{(1)}|-_{\shpp}\ket\propto|\tilde{0}\tilde{0}\tilde{1}\ket-|\tilde{0}\tilde{1}\tilde{1}\ket$
then does not overlap with $|+_{\shpp}\ket=|+_{\sh}\ket$. While such
errors cannot be fully detected as $\bra-_{\cc}|\tilde{\s}_{-}^{(i)}|+_{\cc}\ket\neq0$
for both versions of the code, it is possible to capture the enhancement
with an analogous ``two-level'' quantum recovery developed for those
single-mode codes.

Similarly, we can modify the logical codewords of Steane code by altering
the coefficient of the $|1111111\ket$ computational basis state from
$+1$ to $-1$:
\begin{equation}
\begin{cases}
|0_{\st}\ket & =\frac{1}{\sqrt{8}}\left(|0000000\ket+|1010101\ket+|0110011\ket+|1100110\ket+|0001111\ket+|1011010\ket+|0111100\ket+|1101001\ket\right)\\
|1_{\st}\ket & =\frac{1}{\sqrt{8}}\left(-|1111111\ket+|0101010\ket+|1001100\ket+|0011001\ket+|1110000\ket+|0100101\ket+|1000011\ket+|0010110\ket\right)
\end{cases}\,.
\end{equation}
Just like the original Steane code, $\st$ can provably correct arbitrary
single-qubit Pauli errors. Since a larger set of weight-three errors
is now not completely detectable, $\st$ is a worse-performing code
against local Pauli noise. However, it is slightly better-performing
when it comes to qubit amplitude damping.

Both the Steane code and $\st$ correct one amplitude damping error
and fail to correct two losses in the same way. A difference occurs
when it comes to three-loss errors $\s_{-}^{(i)}\s_{-}^{(j)}\s_{-}^{(k)}$:
Such errors become detectable for $\st$ as, for example, $\s_{-}^{(2)}\s_{-}^{(4)}\s_{-}^{(6)}|1_{\st}\ket\propto-|1010101\ket+|0000000\ket$,
which is orthogonal to $|0_{\st}\ket$. The same occurs for $\{i,j,k\}=\{1,4,5\}$,
$\{3,4,7\}$, and five other combinations corresponding to the five
remaining basis elements of $|1_{\st}\ket$. The remaining possible
loss triples annihilate both $|0_{\st}\ket$ and $|1_{\st}\ket$.
Therefore, all loss triples are detectable. However, the modification
does not improve the code against two-loss errors, and we observe
that the channel fidelity (calculated using a numerical optimal recovery)
improves at most in the third decimal place.

\section{Generic Kerr Hamiltonian for encoding\label{subsec:optimal Kerr}}

\begin{figure}[tb]
\begin{centering}
\includegraphics[scale=0.5]{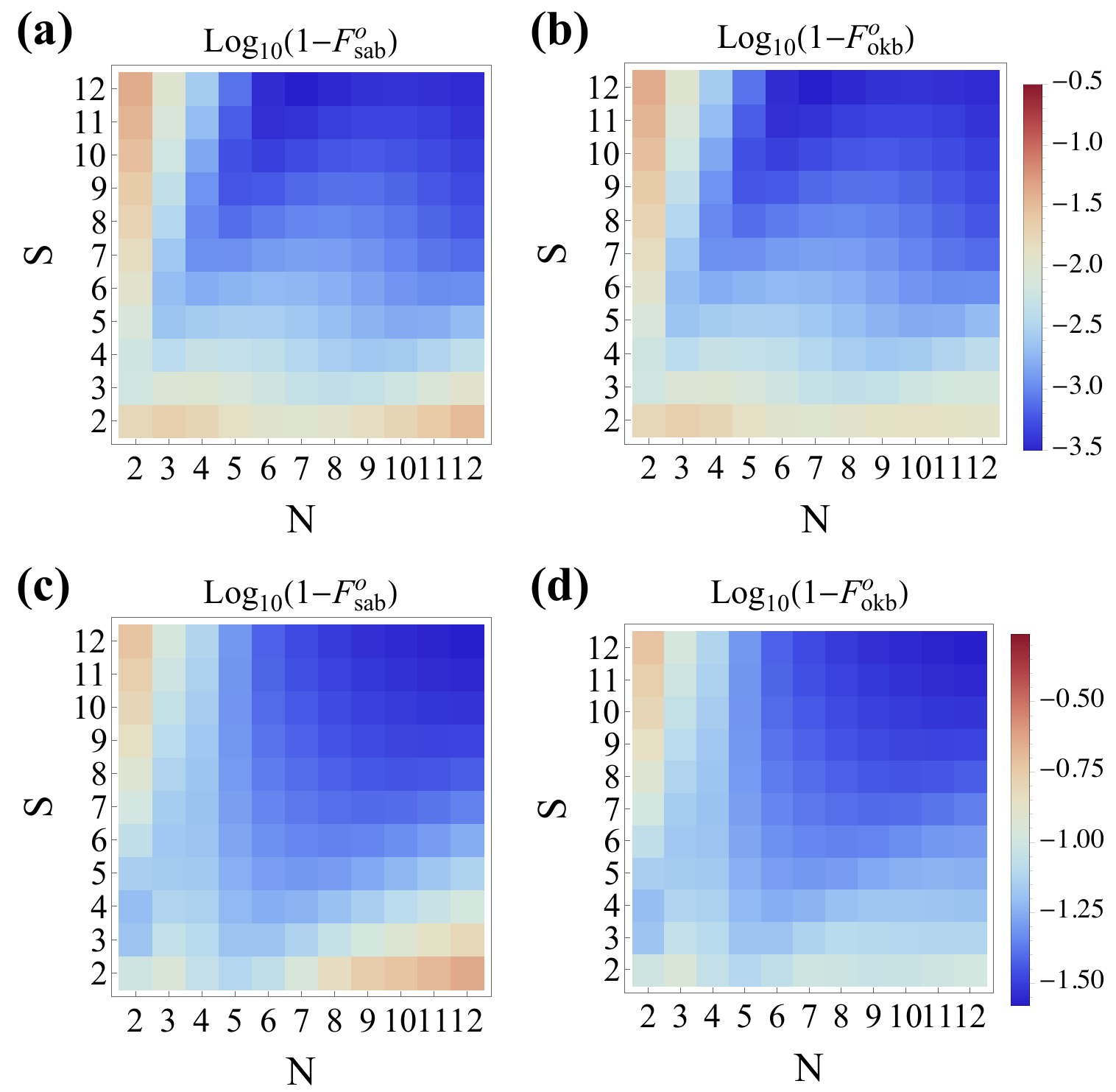}
\par\end{centering}
\caption{Channel infidelities (in logarithmic scale) from optimal recovery
for $\protect\sab$ and $\protect\okb$ codes, respectively, at (a-b)
$\gamma=0.1$ and (c-d) $\gamma=0.25$. Each point represents a code
with associated $S$ and $N$. Please note the color scale is changed.\label{fig:Optimally infidelity_sab v.s. okb}}
\end{figure}

Given $N$, $S$ and $K$, the encoding isometry $\hat{U}_{K}\mathcal{S}_{\bin}(\cdot)\hat{U}_{K}^{\dagger}$
defines a new variant of $\bin$ under Kerr. Dependent on $\gamma$,
one can scan over $K$ to numerically maximize $F^{\mathrm{o}}$ of
the new code. We denote the variant that generates the highest channel
fidelity as optimal-Kerr binomial ($\okb$) code. In Fig.~\ref{fig:Optimally infidelity_sab v.s. okb},
we compare $\log_{10}(1-F_{\sab}^{\mathrm{o}})$ and $\log_{10}(1-F_{\okb}^{\mathrm{o}})$
at $\gamma=0.1$ and $\gamma=0.25$, respectively. We can see that
the difference emerges in the $N\gg S$ region as $\gamma$ gets larger.
In that region, each code word is spanned by a large number of Fock
states (proportional to $N$), and a simple SA becomes suboptimal
for creating destructive interference between error words.
We note that a similar preprocessing -- Gaussian presqueezing --
has been found to improve the fidelity of pure non-Gaussian states
after lossy transmission (\citet{Filip2013,LeJeannic2018}). 

\section{Optimal bosonic codes for excitation loss and dephasing channel\label{subsec:Opt-enc-schs}}

\begin{figure}[!t]
\begin{centering}
\includegraphics[scale=0.5]{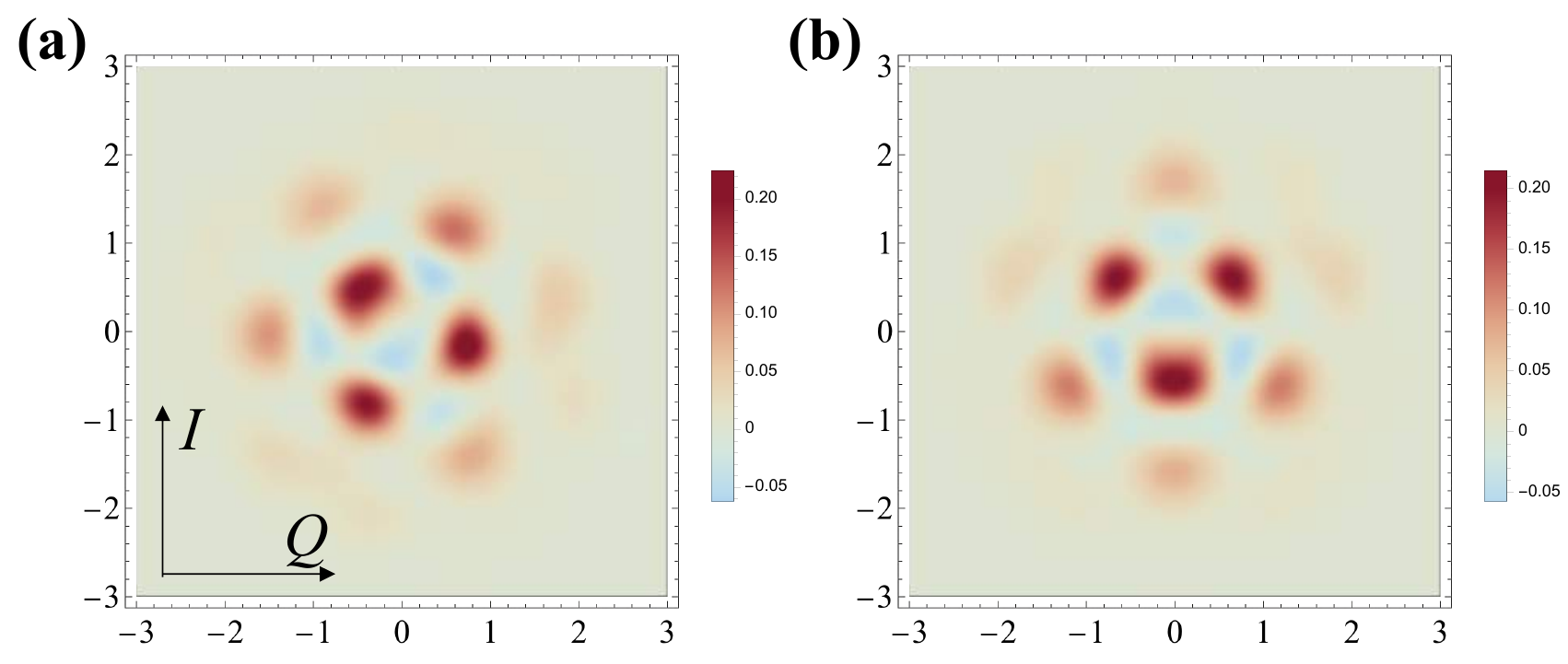}
\par\end{centering}
\caption{Optimal codes from biconvex optimization under energy constraint $\bar{n}_{\mathrm{c}}=2$
for excitation loss channel. (a) Optimal CCQC for excitation loss
channel with $\gamma=0.1$. (b) The CCQC in (a) after a phase rotation
that reduces it to a RCQC. \label{fig: opt codes for excitation loss}}
\end{figure}

We consider the optimal bosonic codes for excitation loss channel.
Since the excitation loss channel $\mathcal{N}_{\gamma}$, with Kraus
operator $\hat{E}_{k}=\sqrt{\frac{\gamma^{k}}{k!}}(1-\gamma)^{\frac{\hat{n}}{2}}\hat{a}^{k}$,
commutes with phase rotation $V_{\theta}=e^{i\theta\hat{n}}$
\begin{equation}
\mathcal{N}_{\gamma}(V_{\theta}\rho V_{\theta}^{\dagger})=V_{\theta}\mathcal{N}_{\gamma}(\rho)V_{\theta}^{\dagger}\,,
\end{equation}
one can apply any phase rotation to $\frac{1}{2}\hat{P}_{\cc}$, which
fully characterizes the encoding subspace, and obtain a new code with
the same QEC performance as $V_{\theta}$ can always be absorbed into
recovery $\mathcal{R}$. In Fig.~\ref{fig: opt codes for excitation loss}a,
we plot the Wigner functions for the maximally mixed state $\frac{1}{2}\hat{P}_{\cc}$
of the optimal CCQC for excitation loss with $\gamma=0.1$ and $\bar{n}_{\mathrm{c}}=2$.
We find that, with a proper $V_{\theta}$, as shown in Fig.~\ref{fig: opt codes for excitation loss}b,
one can obtain a new code with Wigner function of $\frac{1}{2}\hat{P}_{\cc}$
symmetric against Q-axis, i.e. all density matrix entries are real.
The new code is an RCQC, indicating the redundancy of nontrivial phases
in optimal code constructs for excitation loss.
We note that this redundancy of phase does not seem to be lifted as
we consider qudit ($d\geq3$) codes. In Fig.~5a in \citet{Noh2018},
Wigner functions of $\frac{1}{d}\hat{P}_{\cc}$ of optimized qudit
codes ($2\leq d\leq5$) for the excitation loss channel with $\gamma=0.1$
are computed. For all $d$'s considered, the Wigner functions are
in hexagonal lattice which, upon proper phase rotations, will become
symmetric against Q-axis and hence represent a RCQC. These findings
indicate that the necessity of phase in optimal code constructs solely
depends on the nature of channel, not the dimension of encoding.

\bibliographystyle{apsrev4-1}
\bibliography{13_Users_LLS_Dropbox_Research_Constructing_good___ructing_bosonic_codes_via_phase_engineering}